\documentclass[12pt,preprint]{aastex}

\begin{document}

\title{New Solar Seismic Models and the Neutrino Puzzle}

\author{S. Couvidat$^1$, S. Turck-Chi\`eze$^1$, A. G. Kosovichev$^2$}

\email{couvidat@cea.fr, cturck@cea.fr, sasha@quake.Stanford.EDU}

\affil{$^1$ CEA/DSM/DAPNIA/SAp, CE Saclay, 91191 Gif sur Yvette, France}
\affil{$^2$ W.W. Hansen Experimental Physics Laboratory, Stanford 
University, Stanford, CA 94305-4085, USA}

\received{updated 19/02/2002}

\shorttitle{Solar Seismic Models}
\shortauthors{Couvidat, Turck-Chi\`eze, \& Kosovichev}

\begin{abstract}
The SoHO spacecraft made astrophysicists achieve a major breakthrough in the 
knowledge of the Sun. In helioseismology, both GOLF and MDI experiments aboard 
SoHO greatly improve the accuracy of seismic data. More specifically, the detection 
of an enhanced number of low degree low order modes improves the accuracy on 
the sound speed and density profiles in the solar 
core. After a description of these profiles, we build solar seismic 
models. Different models are considered and enable us to derive precise emitted neutrino 
fluxes. These ones are validated by the seismic data and are in agreement 
with the recent detected neutrinos, assuming 3 neutrino flavors. The seismic models are also used to put 
limits on large scale magnetic fields in the solar interior. 
This analysis puts some upper bounds of $\simeq 3\times 10^{7}$ G 
in the radiative zone. Such a field could slightly improve the emitted neutrino flux, 
which remains in agreement with the Sudbury Neutrino Observatory result of 2001.
From the models we deduce gravity mode predictions, and the electron and neutron radial densities that are useful to calculate the 
neutrino oscillations. We also begin to discuss how the external magnetic field may influence such quantities.
\end{abstract}


\section{INTRODUCTION}

Since the detection of oscillations at the surface of the Sun in the early 
sixties (Leighton, Noyes \& Simon 1962), and their later interpretation as 
trapped acoustic waves, helioseismology turned out to be a very powerful tool to 
probe the solar interior. 
Compared to ground networks, the Solar and Heliospheric Observatory spacecraft 
(SoHO) allows very long and continuous measurements leading to very low 
amplitude detections (down to $3 \ \mathrm{mm \, s^{-1}}$ after $4$ years): we 
detect the low frequency part of the oscillation spectrum, not polluted by the 
turbulent surface. We also gained confidence in the frequency measurements from 
comparisons between both the Global Oscillation at Low Frequency (GOLF, see 
Gabriel \mbox{et \mbox{al.}} 1995) and Michelson Doppler Interferometer 
(SOI/MDI, see Scherrer et \mbox{al.} 1995) instruments aboard SoHO: these 
comparisons pointed out the consistency between the two data sets for modes 
between 1.4 and 3.7 mHz (Bertello et \mbox{al.} 2000a).
The SoHO data are of high quality, since roughly speaking the frequencies 
measured with GOLF are affected by an error one order of magnitude smaller than 
the one of 1995. Moreover, the data quality was enforced by progresses in the 
spectral analysis methods. For instance, Bertello et \mbox{al.} (2000a) use the 
new RLSCSA method, while Garc\'\i a et \mbox{al.} (2001) use probability 
computations and different spectral methods (periodogram, random-lag average 
cross-spectrum, homomorphic deconvolution...).
The improvement of seismic data, the addition of new low degree low order p 
modes (Bertello et \mbox{al.} 2000b; Garc\'\i a et \mbox{al.} 2001), and the 
comparison between independent instruments make us confident in the results of 
the inversions based on the GOLF and SOI/MDI data.

Therefore, we utilize these inversions to build solar seismic models. Usually, 
the term ``seismic models'' refers to models that have been directly derived 
from seismic data (\mbox{e.g.} Kosovichev \& Fedorova 1991; Dziembowski et 
\mbox{al.} 1995; Basu \& Thompson 1996; Shibahashi \& Takata 1996; Antia \& 
Chitre 1998). These models use the primary inversion of seismic data that 
returns the sound speed $c_{s}(r)$ and the density $\rho(r)$ profiles inside the 
Sun. By assuming hydrostatic equilibrium, it is straightforward to derive the 
pressure profile $P(r)$. With additional conditions on the input physics, the 
temperature $T(r)$ and helium abundance $Y(r)$ profiles can be determined. The 
knowledge of these different quantities as a function of the fractional radius 
defines a ``seismic model''.\\
In our case this term refers to something different: we stay in the classical 
framework of stellar evolution and we use a 1D stellar evolution code to compute 
a model in agreement with the seismic observations. This agreement is obtained 
by varying a few physical parameters inside their error bars. This is possible 
now, thanks to $10$ years of improvements in solar modelling with the introduction of updated 
physics, microscopic diffusion, \& turbulence at the base of the convection zone 
(e.g. Turck-Chi\`eze et al. 1988; Turck-Chi\`eze \& Lopes 1993; Dzitko et \mbox{al.} 1995; Brun, Turck-Chi\`eze, \& Morel 1998; Brun, Turck-Chi\`eze, \& Zahn 1999). 

The seismic models detailed in this paper are used to derive the neutrino fluxes 
emitted by the Sun. The main interest is that these fluxes include observational 
data, meaning that helioseismology can help in properly determining the neutrino 
production: a key point for the neutrino puzzle.\\
We have presented this first result in Turck-Chi\`eze et \mbox{al.} (2001b) 
showing a remarkable agreement with the detections of the Sudbury Neutrino 
Observatory (hereafter SNO) plus SuperKamiokande (SK) experiments. In 
this paper, we detail and generalize our approach.
In section 2 we introduce the data we use and the major features of the 
evolution code that permits us to compute models. In section 3 we discuss the 
way the sound speed and density inversions are carried out. In section 4 we 
derive the seismic models and discuss about their uniqueness. In section 5 we compute 
the seismic oscillation frequencies. In section 6 we compare the predicted neutrino fluxes to the ones detected by the terrestrial experiments like SK or SNO, and derive some quantities useful to the 
determination of the neutrino oscillation parameters.  In section 7 we focus on the magnetic 
field and the way we can put upper bounds on it. Finally we conclude in section 
8.

\section{THE HELIOSEISMIC OBSERVATIONS AND THE PHYSICAL INPUTS OF THE SOLAR MODELS}

To derive the sound speed ($c_{s}$) and density ($\rho$) of the Sun, we use the 
data from both the GOLF and MDI instruments and the inversion procedure described 
in the next section.
For the p modes with $\ell \le 2$, we use Bertello et \mbox{al.} (2000a, 2000b). 
These frequencies are extracted from GOLF data.
For the p modes with $\ell \ge 3$, we use Rhodes et \mbox{al.} (1997) who 
utilized data from the SOI/MDI instrument.
The quality of these modes has been discussed in Turck-Chi\`eze et al. (2001b)
showing the results of some inversions with different values of the radial order $n$; in 
section 5 we show, by comparing observed and calculated frequencies, 
how the SoHO observations have improved the quality of the seismic data. 

The solar models are computed with the CESAM code (Code d'Evolution Stellaire 
Adaptatif et Modulaire, see Morel 1997). This is a 1D quasi-static stellar evolution 
code that solves the stellar structure equations by a spline collocation method. 
We always start the evolution from the pre-main sequence (PMS). The basic 
physical characteristics of the models are:

\begin{itemize}
\item the nuclear reaction rates from Adelberger et \mbox{al.} (1998), with 
Mitler intermediate screening (Mitler 1977). For the 
$\mathrm{^{7}Be(p,\gamma)^{8}B}$ reaction, we use the $S(0)_{17}$ value derived 
by Hammache et \mbox{al.} (1998). For the $\mathrm{^{7}Li(p,^{4}He)^{4}He}$ 
astrophysical factor we use Engstler et \mbox{al.} (1992). In order to properly 
compute the lithium burning on PMS, we adjust the time step and the rotation 
law, according to Piau \& Turck-Chi\`eze (2001);
\item the opacities are derived from the OPAL95 opacity tables (Iglesias \& 
Rogers 1996) for temperatures larger than $5600 K$. For lower temperatures, we 
use the Alexander opacities  (Alexander \& Ferguson 1994);
\item the equation of state (EOS) is OPAL (Rogers \& Iglesias 1996);
\item the microscopic diffusion is taken into account with the prescription of 
Michaud \& Proffitt (1993);
\item the turbulent mixing at the base of the convection zone (hereafter BCZ) is 
treated, following Brun et \mbox{al.} (1999).
\end{itemize}

The references listed emphasize the improvement in the solar interior physics 
that occured in the last decade, in parallel to the improvement of the seismic data. 
This allowed us to reject some extra physical processes like large screening 
and mixing in the central region (\mbox{e.g.} Turck-Chi\`eze et al. 2001a).

\section{SEISMIC DATA AND THE INVERSION OF THE SOLAR SOUND SPEED AND DENSITY}

The inversion results for the sound-speed and density profiles
are obtained by using the Optimally Localized Averaging (OLA)
method (see Kosovichev 1999 ---hereafter K1999--- for mathematical details).
In this method the differencies between
the observed and model frequencies are expressed as
a sum of two linear integrals for the corresponding relative differences
in the sound speed and density (see \mbox{eq.} [52] of K1999). The sensitivity kernels in these
integrals are calculated by using a variational principle for
adiabatic non-radial oscillations that allows us to neglect the variations of the eigenfunctions to
the first order of approximation. 
The non adiabatic effects, and the uncertainties in the physics of the near-surface layers, are taken into account by adding a term to the two integrals related to the frequency differences.
This term is a smooth function of frequency, scaled with the mode inertia (see \mbox{eq.} [72] of K1999).

The estimates of the localized averages
for the sound-speed and density corrections to the solar models are obtained by considering linear combinations
of the integral relations for the frequency differences.
We proceed such that the corresponding linear
combinations of the sensitivity kernels form narrow localized,
Gaussian-type, kernels at various target positions along the solar
radius for one of the variables (sound speed or density) and
are negligible for the other variable. The inversion
procedure includes additional constraints to eliminate the
surface term approximated by a linear combination of Legendre
polynomials of degree less than 5, and also
to minimize the errors of the sound-speed and density
corrections. The later constraint is based on observational error estimates
of the mode frequencies, and includes a regularization parameter that
controls the trade-off between the spatial resolution of the inversions
(measured as a ``spread'' of the averaging kernels) and the error
magnification. The regularization parameter is chosen to provide
a sufficiently smooth radial dependence of the sound-speed and density
corrections.

\begin{figure*}[htb]
\plotone{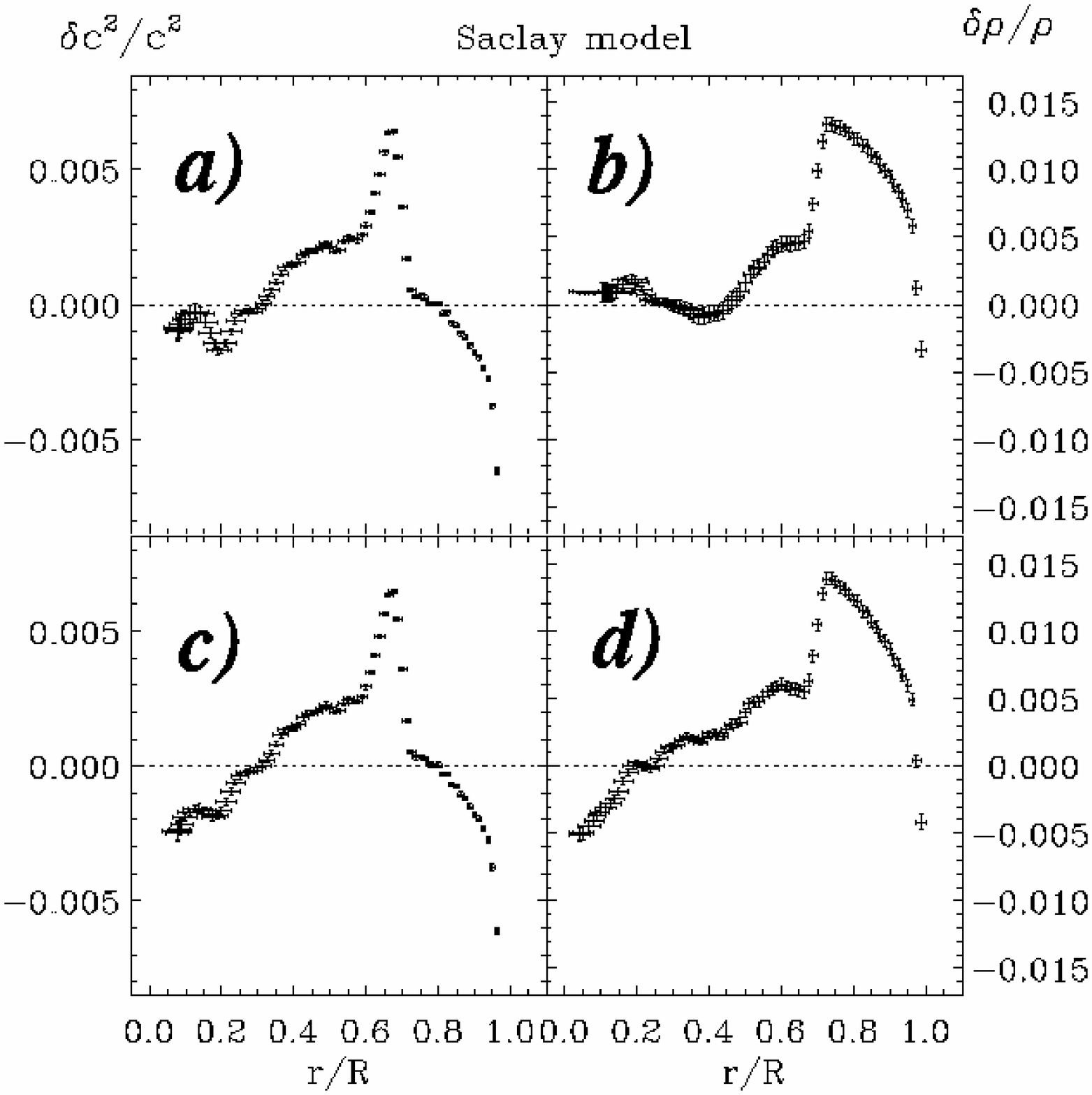}
\caption{\label{inversion} {\it Sound speed and density inversion using a
 standard Saclay model (Brun et \mbox{al.} 1999). In these plots, panels a) and b)
are obtained from the GOLF modes of l=0-2 above 1 mHz (Bertello et al. 2000a,b) and MDI medium-l
data (l=3-250); the total number of modes is 2183. 
Panels c) and d) are obtained by adding three modes provided by Bertello et al. (2000b)(l=0, n=3,5,6)
to the previous combination of p modes.} }
\end{figure*}

The results are presented in our figures in the form of the
horizontal and vertical bars centered at the center of gravity
of the localized averaging kernels. The size of the horizontal
bars corresponds to a characteristic width (``spread'') of the
averaging kernels and provides an estimate of the spatial
resolution, and the vertical bars correspond to 1$\sigma$ formal
error of the sound-speed and density corrections (see Table 1 of Turck-Chi\`eze et al. 2001b).
The standard solar model of Brun et \mbox{al.} (1999) was used as
a reference. The inversion procedure was tested by using
various other solar models as Sun's proxy, and adding random
Gaussian noise to calculated frequencies of these models
to simulate the observational errors.

Figure 1 shows the inversions obtained for the sound speed and the density. This figure highlights 
how the low order 
modes impact deeply on the density profile and how they slightly improve the sound 
speed in the solar core. In the following, we shall call seismic model a model that reproduces as well
as possible the observed sound speed profile. We will not take the density as
a reference yet, since we first wait for confirmation of its profile by new detections at low frequencies,
 due to its extreme sensitivity to these modes.
The density appears more sensitive to the detailed physics than the sound speed. 
It will be used in the future to progress on the dynamics of the Sun.

\section{BUILDING THE SEISMIC MODELS}

\subsection{Starting Point}

\begin{figure*}[htb]
\epsscale{0.8}
\plotone{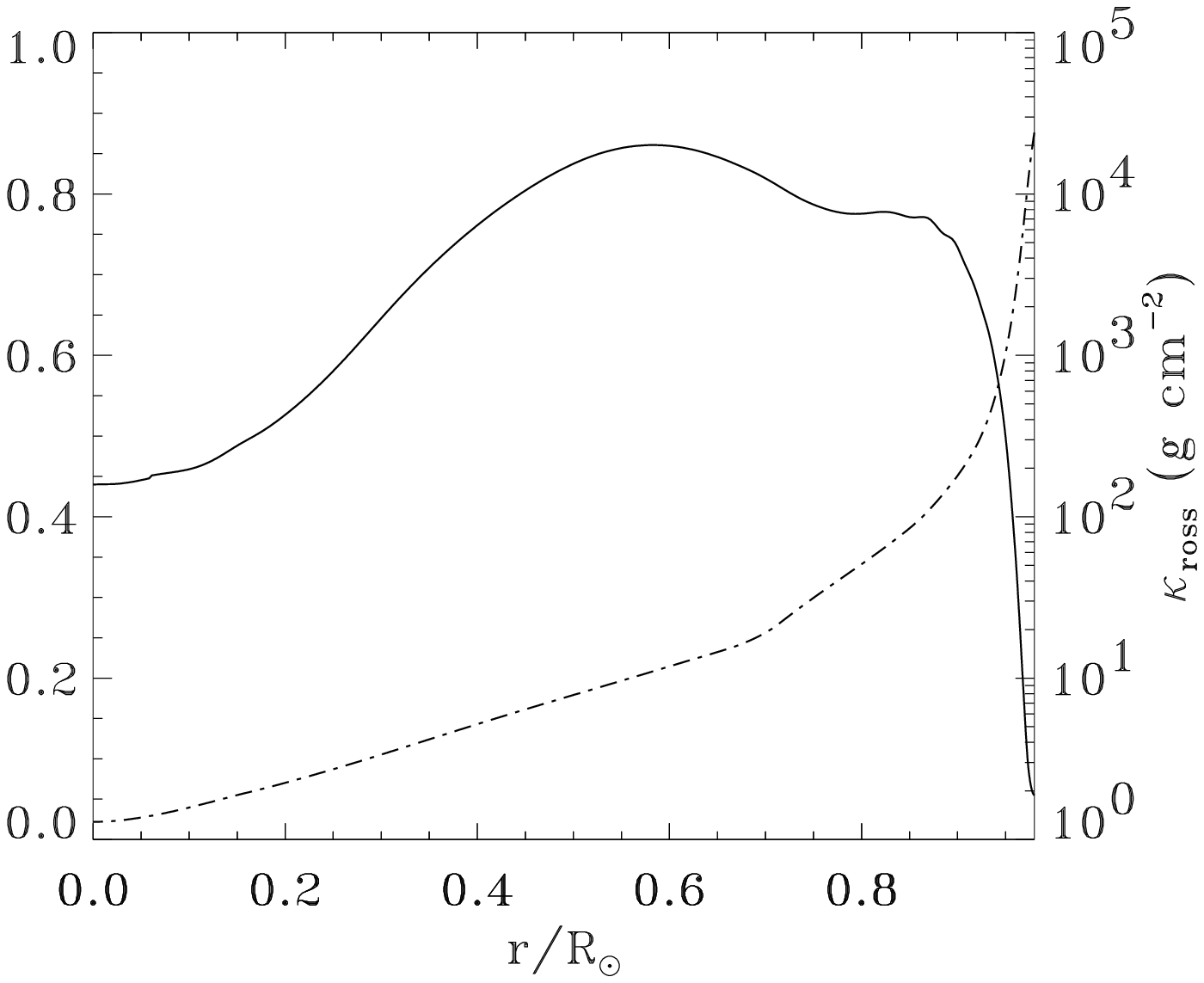}
\plotone{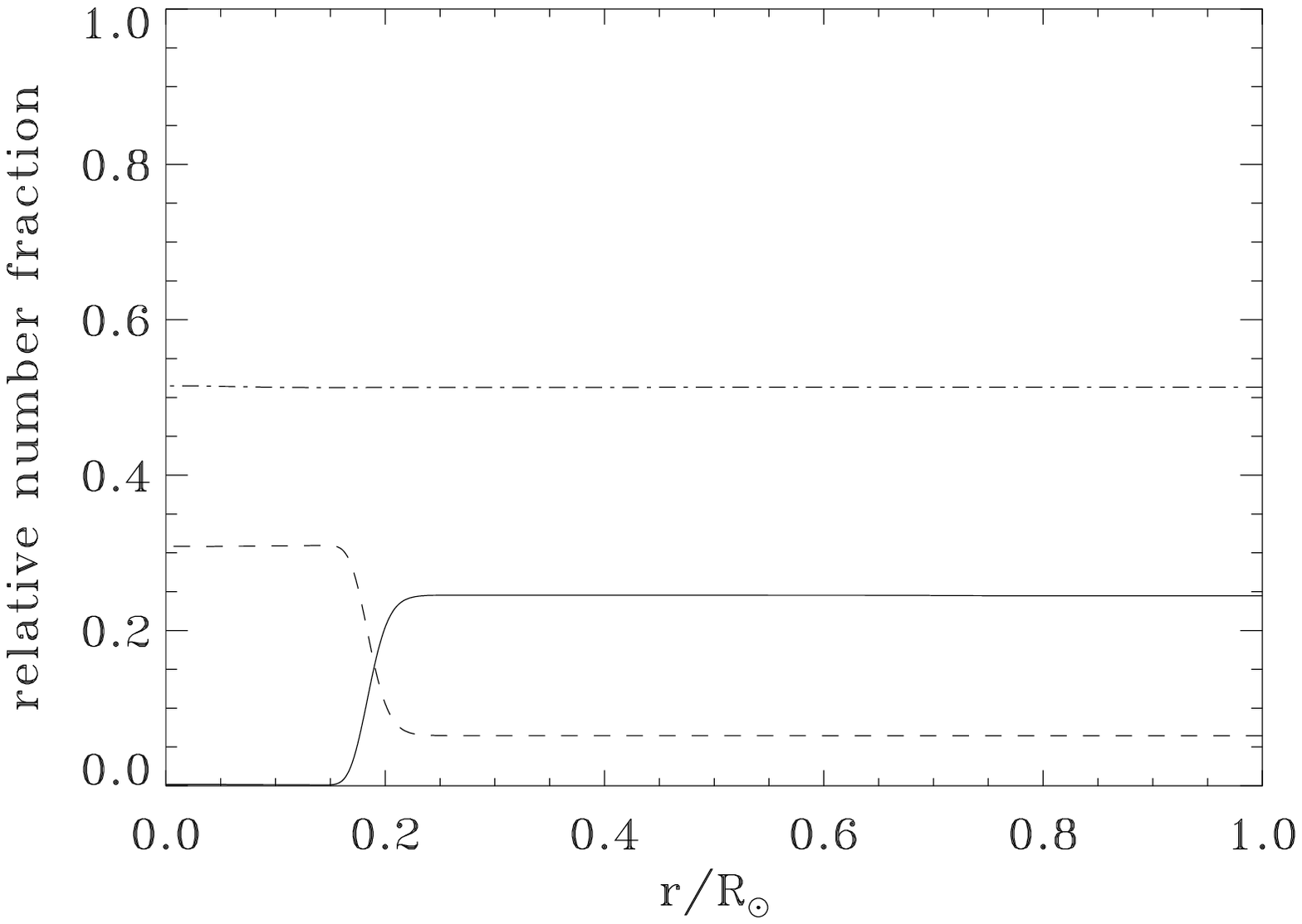}
\caption{\label{opaabond} {\it Upper figure: the contribution of the heavy elements to the 
opacity (plain curve). Superimposed is the profile of the Rosseland opacities 
$\kappa_{ross}$ (dot-dashed curve). Lower figure: number fraction of three elements related to the 
metal number fraction: the carbon (plain curve), the nitrogen (dashed curve), 
and the oxygen (dot-dashed curve). Solar composition at $4.6$ Gyr.}}
\end{figure*}

In this study, our goal is to cancel the discrepancy on $c_{s}$ between the 
solar models and the real Sun. To this end, we slightly modify a few physical 
parameters used in CESAM. Of course, we derive seismic models that are not 
unique: we could obtain similar results on the sound speed by adjusting 
different parameters. However, we justify our choices in section 4.3. We adjust 
the physical quantities the sound speed is sensitive enough to, and modify them 
within their error bars. It is decided to change as little parameters as 
possible.

We start from a specific solar model: the non-standard Btz model of Brun et \mbox{al.} (1999). This model was designed to reduce the 
discrepancy on $c_{s}$ between the standard models and the Sun at the BCZ, by 
taking into account the horizontal motion produced by the sudden disappearance 
of the differential rotation profile in the tachocline. Doing so, they get for 
the first time a correct $\mathrm{^{7}Li}$ abundance at the solar 
surface ($A_{\mathrm{^{7}Li}} = 1.16\pm0.10$ dex according to Grevesse \& Noels 
1993 [hereafter G\&N93]): the only abundance that was not predictable before.
It is a real improvement since the standard solar models do not consider any 
turbulent mixing at the BCZ. It is well known that the lithium is the only indicator 
of the internal structure for numerous stars and its abundance is very difficult 
to predict in classical stellar evolution. However, the adopted tachocline prescription is 
purely hydrodynamic and does not account for any magnetic field, eventhough the 
magnetic dynamo process is thought to occur in the thin tachocline.

The main impacts of this prescription are: to reduce the influence of the 
microscopic diffusion ---the diffusion of heavy elements toward the center of 
the Sun is slowed down--- and to burn $\mathrm{^{7}Li}$ on the main sequence. 
Three parameters define the tachocline:
\begin{itemize}
\item its current width, taken as $d = 0.05 R_{\odot}$;
\item the Br\"unt-V\"ais\"al\"a frequency at the BCZ, $N = 25 \, \mu$Hz;
\item the present rotation rate at the BCZ, $\Omega_{0} = 415$ nHz.
\end{itemize}
Moreover, the turbulent diffusion coefficient in the tachocline is 
time-dependent, since it is related to the rotation rate. In the Btz model the 
rotation as a function of time follows the Skumanich's law (Skumanich 1972).

\subsection{The First Seismic Model: Seismic$_{1}$}

Starting from Btz, we undertake the construction of a first seismic model 
(already presented in Turck-Chi\`eze et al. 2001b). We focus primarily 
on the solar core where the energy generation and neutrino production occur: to 
adjust the physical parameters, we take advantage of our knowledge of the 
sensitivity of the model to the physical ingredients (through, for instance, 
Turck-Chi\`eze \& Lopes 1993; Dzitko et \mbox{al.} 1995; Turck-Chi\`eze et 
\mbox{al.} 1997; Turck-Chi\`eze et \mbox{al.} 2001a).

\begin{figure*}[htb]
\epsscale{0.8}
\plotone{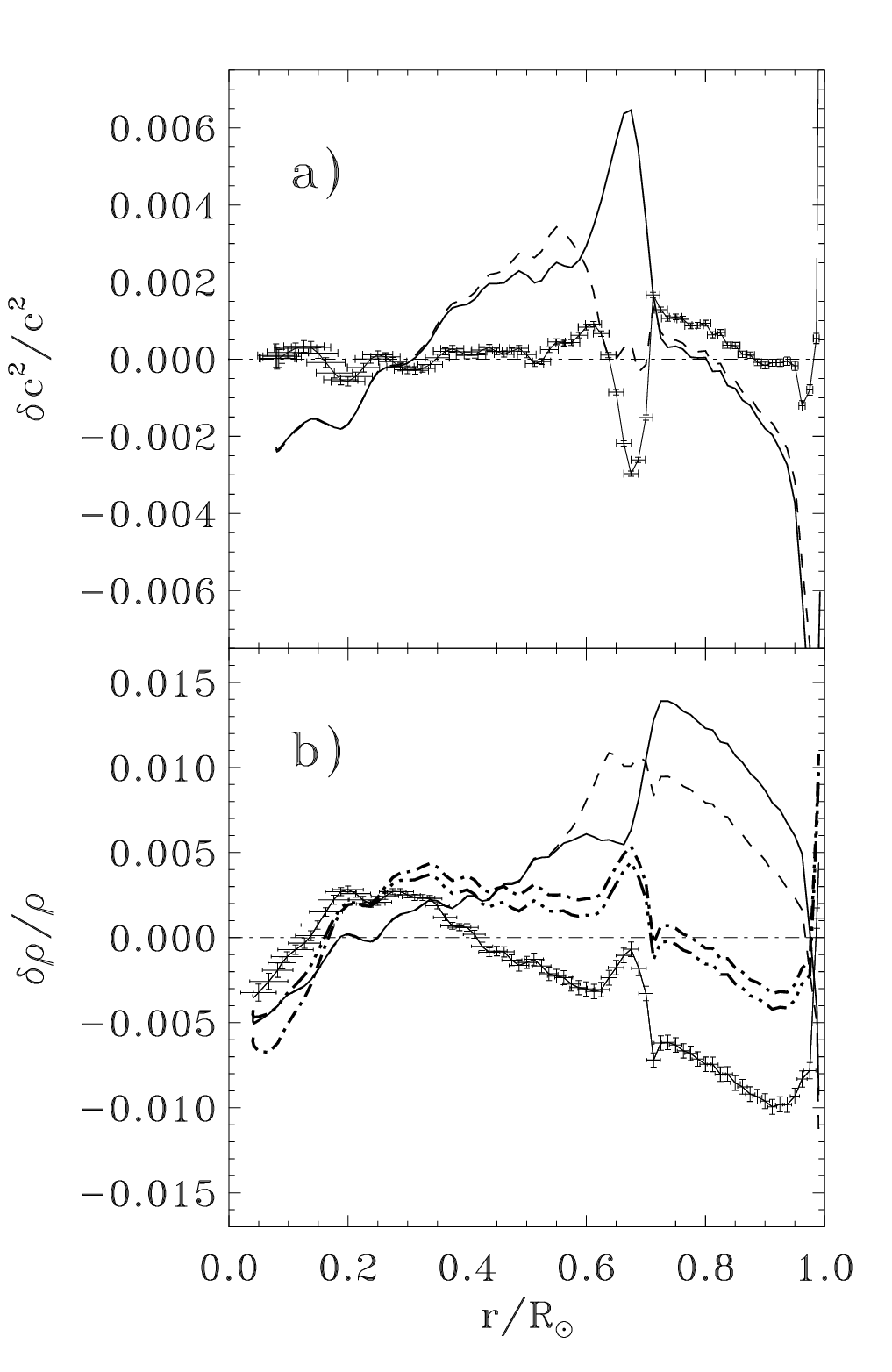}
\caption{\label{sismique1} {\it Seismic$_{1}$ model (plain curve with error 
bars) : a) difference in the square of the sound speed between the Sun and the 
model; b) difference in the density between the Sun and the model.
The plain curves with no error bar are for the Saclay standard solar model, 
while the dashed curves are for the Btz model.
Two other models based on seismic$_{1}$ are considered on the density 
figure: the model with the $^{3}He(^{4}He,\gamma)^{7}Be$ reaction rate reduced 
by $10\%$ (dot-dot-dot-dashed curve) and the model with the CNO poly-cycle 
reaction rates reduced by $70\%$ (dot-dashed curve) [From Turck-Chi\`eze et al. 
2001b].}}
\end{figure*}

We increase the $p-p$ cross section (hereafter $S(0)_{p-p}$) by $1\%$: usually 
$S(0)_{p-p}= 4.00(1 \pm 0.007^{+0.020}_{-0.011}) 10^{-22}$ keV b (Adelberger et 
\mbox{al.} 1998). As this cross section is only known theoretically, we try to 
constrain it by helioseismology. The change of $S(0)_{p-p}$ has a rather large 
impact on $c_{s}$. An increase of this cross section induces a decrease of the core temperature, since the model 
is calibrated to obtain the solar luminosity.

In parallel to this modification, we adjust the initial metallicity 
($Z_{0}$) of the Sun. Compared to Btz ($Z_{0}=0.01959$), we increase it by 
$3.9\%$. In the radiative interior, a major contribution to the opacities comes
from the heavy elements (see Fig. \ref{opaabond}): at $0.6 R_{\odot}$, about $85 
\%$ of the opacities are due to the metals. In the solar core, this part is 
still $45 \%$, in this case all the elements are almost fully ionised, except 
iron. Actually, the opacities in the core are primarily due to inverse 
brehmstrahlung and electron scattering, while at $0.6 R_{\odot}$ the bound-bound 
processes dominate. By increasing $Z_{0}$, we change both the Rosseland opacities 
($\kappa_{ross}$)  and the mean molecular weight ($\mu$), explaining why the 
sound speed of the seismic model decreases in the core and goes up in the 
radiative zone: in the core, the raise of $\mu$ dominates the raise in 
temperature produced by the increase of $\kappa_{ross}$; beyond $0.3 R_{\odot}$, 
the raise in temperature overcomes the increase of $\mu$.
 A last change is carried out: the OPAL EOS is tabulated and depends on $Z_{0}$. 
Instead of $Z_{0} = 0.01959$ used to compute the EOS table for Btz, we use a 
table computed with $Z_{0}=0.0203$. Actually, this change has no important 
impact on the sound speed profile.

Having reduced the discrepancy on $c_{s}$ in the radiative zone, we undertake to 
diminish it at the BCZ. To this end we modify the parameters defining the 
tachocline: we reduce its width from $0.05 R_{\odot}$ to $0.025R_{\odot}$, in 
accordance with the most recent helioseismic results (\mbox{e.g.} Elliott et 
\mbox{al.} (1998) announce $d=0.02 R_{\odot}$, Corbard et \mbox{al.} (1999) 
announce $d < 0.05 R_{\odot}$); we also increase the present rotation rate 
$\Omega_{0}$ from 415 to 430 nHz (Corbard et \mbox{al.} 1999). 
 Moreover, to retrieve a good $\mathrm{^{7}Li}$ abundance at the surface, we 
increase the Br\"unt-V\"ais\"al\"a frequency $N$ from 25 to 105 $\mu$Hz. This 
frequency undergoes a dramatic change when approaching the convection zone. An 
increase of $N$ reduces the efficiency of the turbulent mixing and diminishes the 
$\mathrm{^{7}Li}$ depletion.

Another improvement is the change in the rotation law. Instead of the 
Skumanich's law, which is far less realistic on the PMS, we utilize the law proposed by 
Bouvier et \mbox{al.} (1997), following Piau \& Turck-Chi\`eze (2001): the 
rotation rate of the Sun slightly goes down during the early evolution phase, and 
experiences a rapid acceleration at $10$ million years when the circumstellar 
accretion disk separates from the young Sun. Finally, the rotation rate 
decreases following the Skumanich's law because of the magnetic breaking. 
Eventhough the impact of such a rotation law on the solar evolution is rather 
weak, its use is more realistic. To reach a correct $\mathrm{^{7}Li}$ content at 
$4.6$ Gyr, we suppose that the Sun was a slow rotator on the PMS (this is why we 
choose a disk separation at $10$ million years).

To complete the building of the seismic model, we concentrate on the outer part 
of the $\delta c_{s}^{2}/c_{s}^{2}$ profile: it is still far from flat, since no 
appropriate model exists for the upper layers with a 1D code. However, it is 
possible to obtain a better agreement with the Sun by calibrating the seismic 
model at a radius $R_{1}$ different from the standard one $R_{\odot} = 6.9599 
\times 10^{10}$ cm deduced from photometric observations (\mbox{e.g.} Allen 
1976). Recent analyses based on f-mode frequencies (e.g. Schou et \mbox{al.} 
1997; Antia 1998) worked out {\it seismic} radii, respectively $R$ = 695.78 Mm and $R$ = 695.68 Mm, slightly smaller than the {\it photometric} one. This latter is greater than the recent
optical determination $R$ = 695.51 Mm by Brown \& Christensen-Dalsgaard (1998).  
There is neither a firm answer explaining the origin of these discrepancies, 
nor the effect of the solar cycle on the radius. Thus, we calibrate the seismic$_{1}$ model with $R_{1}= 6.95936 
\times 10^{10}$ cm, with a consequent {\it ad hoc} improvement in the convective zone.
Nevertheless, the inversion of the density and sound speed was carried out 
assuming the standard value for $R_{\odot}$.\\

The Table \ref{tab1} lists all the main features of the models described in this 
paper.
The  seismic$_{1}$ model is also available on the web site {\it 
http://apc-p7.org/Neutrino\_APC/Sismic\_model.html}, with the detailed values of 
many parameters for a large number of shells (including the electron number 
density, see \mbox{sect.} 6.3).

\subsection{Justification of the Changes for the Seismic$_{1}$ Model}

The adjustments made increase the overall agreement between the Sun and the 
solar model on the $c_{s}$ profile (see Fig. \ref{sismique1}). The improvement 
is less obvious when checking at the density profile but the progress is real 
too. 
To reach this improvement in the core and the radiative zone, we restricted the 
modifications to two parameters, $S(0)_{p-p}$ and $Z_{0}$. With such a 
restriction, it can be shown that we need to increase both these quantities to 
cancel (at least reduce) the discrepancy on $c_{s}$.
 
First, G\&N93 work out $(Z/X)_{s} = 0.0245\pm10\%$ at the solar surface; this 
ratio has an impact on $Z_{0}$, whose value is larger than the present 
photospheric $Z$ value because of the microscopic diffusion. Let us suppose we 
want to cancel the sound speed difference by increasing $Z_{0}$ and reducing (or 
not changing) $S(0)_{p-p}$: to keep a $(Z/X)_{s}$ ratio within its error bar, we 
must limit the raise of $Z_{0}$ to $\approx 5.5\%$. This raise is lower than the 
one needed to cancel $\delta c_{s}^{2}/c_{s}^{2}$. Thus, if we increase $Z_{0}$, 
we also need an increase of $S(0)_{p-p}$ to get $\delta c_{s}^{2}/c_{s}^{2}=0$. 
Second, let us suppose we diminish $Z_{0}$: to cancel the sound speed difference 
in the core, we need to increase $S(0)_{p-p}$ beyond its upper error bar (equal 
to $2.0\%$).

Thus, the error bars on $S(0)_{p-p}$ and on $(Z/X)_{s}$ imply an increase of 
both parameters to cancel the discrepancy on $c_{s}$ below $0.6 R_{\odot}$.

\subsection{Non Uniqueness of the Seismic$_{1}$ Model: Alternative Seismic 
Models}

Of course, the seismic$_{1}$ model is not unique:

First, we could have changed other nuclear cross sections instead of 
$S(0)_{p-p}$. Yet, Turck-Chi\`eze et \mbox{al.} (2001a) highlight the lack of 
sensitivity of the sound speed to the other nuclear reactions, like 
$\mathrm{^{3}He(^{3}He,2p)^{4}He}$, $\mathrm{^{3}He(^{4}He,\gamma)^{7}Be}$ or 
the CNO bi-cycle. For instance, an increase of 25$\%$ for the first two 
reactions mentioned changes $\delta c_{s}^{2}/c_{s}^{2}$ by only $\approx 
0.1-0.2\%$, and thus does not improve significantly the agreement with the Sun. 
On the other side, the 
sound speed is sensitive enough to the $p-p$ reaction rate. Therefore it seems 
more appropriate to only adjust $S(0)_{p-p}$: we build the best model with the 
most simple assumptions. On \mbox{Fig.} \ref{sismique1}, you can see that the density is a little bit more 
sensitive to these reaction rates than the sound speed.

Second, with the accuracy reached on the sound speed, this quantity is sensitive 
to many physical parameters whose uncertainties can be quite large. It is 
primarily sensitive to $S(0)_{p-p}$, the opacities, the heavy element 
abundances, the microscopic diffusion process, but it can also undergo changes 
under ``secondary'' parameters such as the solar age or the solar radius.

\subsubsection{The Seismic$_{2}$ Solar Model}

\begin{figure*}[htb]
\epsscale{0.4}
\plotone{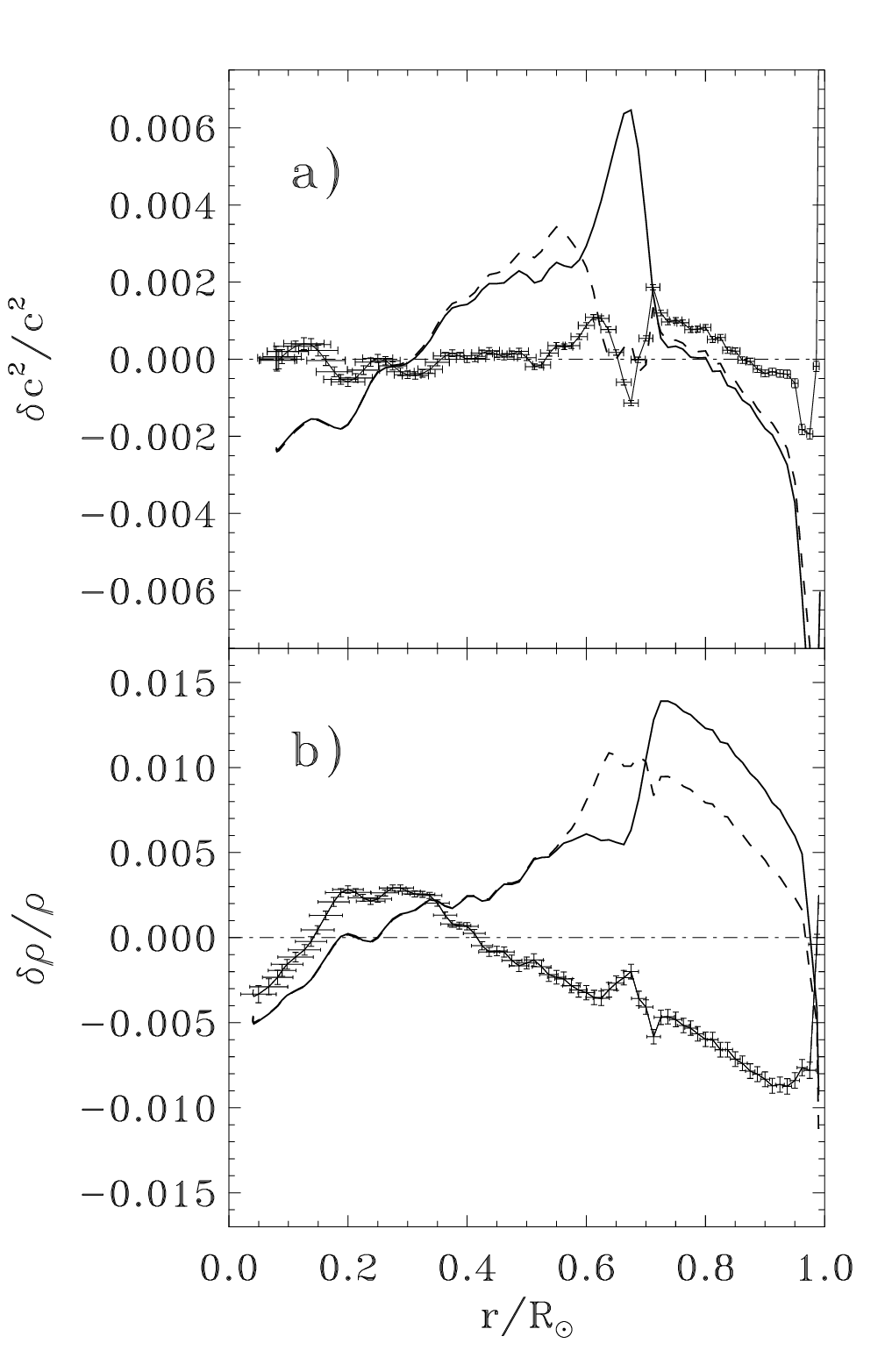}
\plotone{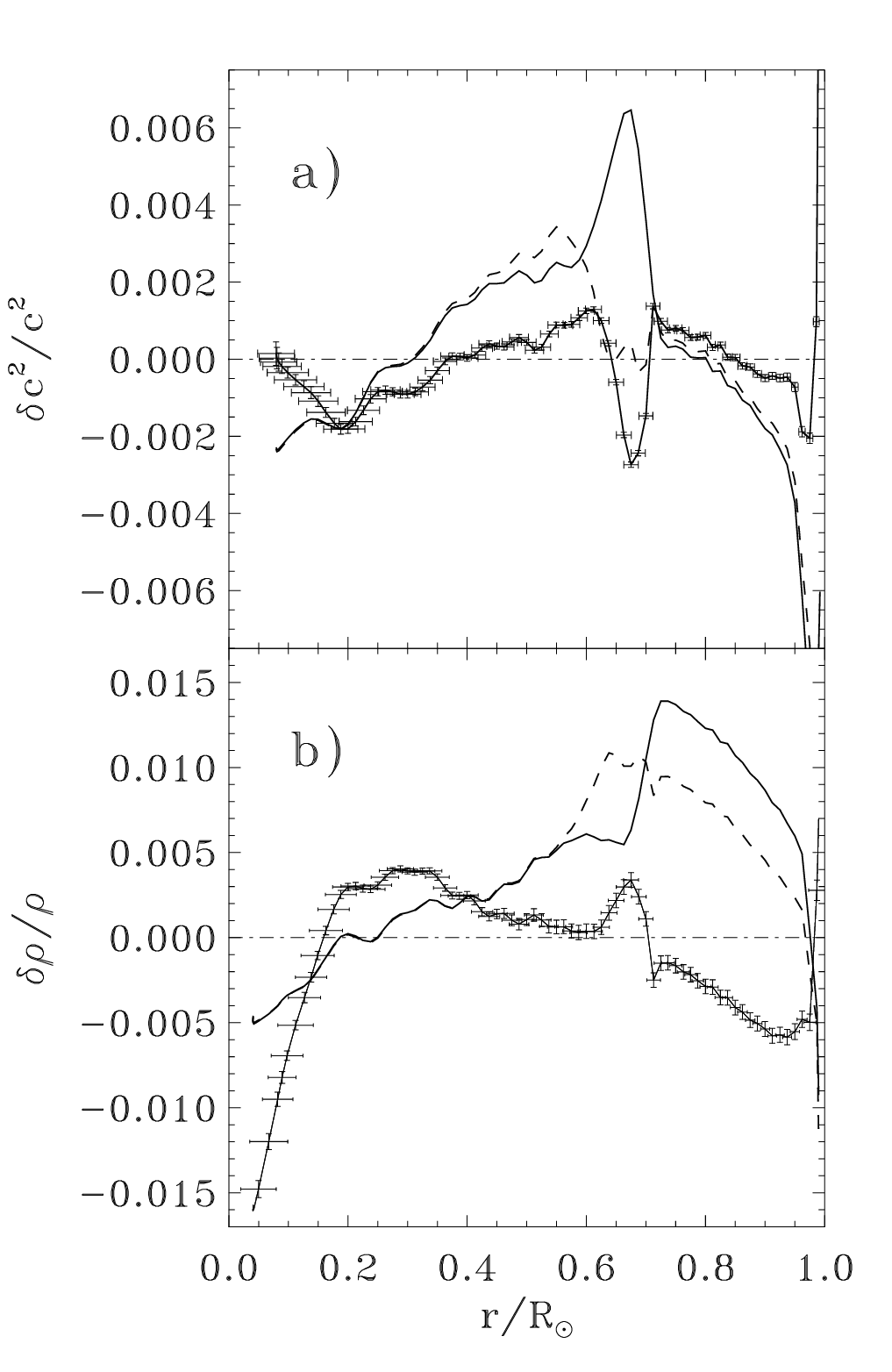}
\caption{\label{sismique23} {\it Same caption as the previous figure 
for the seismic$_{2}$ model (left panels) and the seismic$_{3}$ model (right panels) (see text). }}
\end{figure*}

Actually, a second seismic model is constructed from Btz by modifying the 
opacities $\kappa_{ross}$, reducing $Z_{0}$ by $3.5 \%$ and increasing 
$S(0)_{p-p}$ (we also calibrate this model to $R_{2}= 6.95866 \times 10^{10}$ cm). Its main advantage is to 
reach a $(Z/X)_{s}$ ratio equal to the one proposed by G\&N93, while the one of 
the seismic$_{1}$ model approaches the upper limit of the error bar. By lowering 
$Z_{0}$ we reduce the mean molecular weight in the central part of the Sun, and 
the opacities as well. On the other side, we raise $\kappa_{ross}$ following 
Brun et \mbox{al.} (1998), by simulating an increase of about $7.5\%$ 
in the C, N, and O opacities. This raise could be attributed to uncertainties in 
the bound-bound and bound-free processes, i.e. to an intrinsic error in the 
opacity calculations.
 The change in the opacities of the CNO elements produces an increase of about 
$1.5\%$ in $\kappa_{ross}$ in the solar core, and about $4.5\%$ below the 
convection zone. This raise compensates for the decrease due to the reduction of 
$Z_{0}$, so that we get roughly: no opacity change in the core and an increase 
of about $1.5\%$ at $0.6 R_{\odot}$. We just operate this {\it ad hoc} 
adjustment below $0.6 R_{\odot}$ to diminish the discrepancy on $c_{s}$ at the 
BCZ. As for the first seismic model, we also need an increase of $S(0)_{p-p}$: 
by $1.3\%$.  
This seismic$_{2}$ model produces a sound speed profile  similar to the one of 
the seismic$_{1}$ model, and even better at the BCZ (see \mbox{Fig.} \ref{sismique23}).

However, for the rest of the paper, we just consider the seismic$_{1}$ model to 
derive the neutrino fluxes and to test the magnetic field: this model contains 
no {\it ad hoc} adjustment, less parameters were changed \& the error bars on 
$S(0)_{p-p}$ and $Z_{0}$ (through the $(Z/X)_{s}$ ratio) are more or less well 
known while the uncertainties on $\kappa_{ross}$ remain unknown. However, the 
great effort done by laser measurements and the theoretical successes of the 
Livermore group lead to specific uncertainties on $\kappa_{ross}$ of the order 
of $5\%$.

\subsubsection{The Opacities and the Change in Metal Abundances}

Just a few words about these opacities: the OPAL95 tables we use to interpolate 
the $\kappa_{ross}$ values were calculated assuming a G\&N93 mixture. Along the 
solar evolution the metal composition in the core changes: because of the CNO 
poly-cycle and the microscopic diffusion, the relative number fractions of the 
metals are modified (see \mbox{Fig.} \ref{opaabond} for the composition at $4.6$ 
Gyr). This should have an impact on the opacities: we have to correct them as 
soon as the solar core composition is ``far'' from the G\&N93 mixture (this 
occurs at an age of, roughly, 100 million years). Starting from the 
seismic$_{1}$ model, we compute a very similar model but with a correction for 
$\kappa_{ross}$ below $0.15 \, R_{\odot}$ to take into account the change in the 
composition. Actually, this correction is very small: from $-0.88\%$ at $0 
R_{\odot}$ to $-0.5\%$ at $0.15 R_{\odot}$, for the Sun at $4.6$ Gyr. It has a 
minute impact on the sound speed profile: it just increases $\delta 
c_{s}^{2}/c_{s}^{2}$ by $0.0005$ in the core! However, with the precision we 
reach on $c_{s}$, we can test such a small effect. This latter leads to that, 
instead of increasing $S(0)_{p-p}$ by $1\%$ for the seismic$_{1}$ model, an 
increase by only $0.75\%$ is enough.\\

\subsubsection{The Impact of the Age: Seismic$_{3}$}

Finally, we also check the sensitivity of $c_{s}$ to the solar age 
($t_{\odot}$). A recent work by Dziembowski et \mbox{al.} (1999) used 
helioseismology to determine $t_{\odot}$. Depending on the method they use, they 
derive very different results. However, using the small frequency separation 
which they claim to be the most accurate measure, they conclude that $t_{\odot} 
= 4.66 \pm 0.11$ Gyr (the lifetime on the PMS must be added). 
Starting from Btz, we compute the seismic$_{3}$ model: those are basically 
the same models except that $t_{\odot} = 4.735$ Gyr for the second one, the 
tachocline parameters are changed (as for the previous seismic models), and we 
calibrate at $R_{2}$. The augmentation by only $\approx 2.9\%$ of $t_{\odot}$ 
produces a rather important change in the $c_{s}$ profile and reduces the 
discrepancy with the Sun, eventhough the reduction is less significant than with 
the other seismic models (see \mbox{Fig.} \ref{sismique23}). The seismic$_{3}$ 
model shows that, supposing we under-estimate the solar age, an increase of 
$S(0)_{p-p}$ and $Z_{0}$ less important is needed to retrieve the flat $\delta 
c_{s}^{2}/c_{s}^{2}$ profile of the seismic$_{1}$ model. Eventhough this does 
not challenge the need to raise both $S(0)_{p-p}$ and $Z_{0}$, this reduces the 
amplitude of the increase. On the contrary, a solar age less than $4.6$ Gyr 
would favor a larger increase of the two parameters. It is noticeable that the 
density profile greatly disfavors the seismic$_{3}$ model (see \mbox{Fig.} 
\ref{sismique23}, right figure, panel b), unlike the sound speed profile.

\begin{deluxetable}{lcccccc}
\tablewidth{0pt}
\tablecaption{Main Features of the Solar Models \label{tab1}}
\tablehead{\colhead{} & \colhead{Btz} & \colhead{Seismic$_{1}$} & 
\colhead{Seismic$_{2}$} & \colhead{Seismic$_{3}$} & 
\colhead{Seismic$_{1}$B$_{1}$} & \colhead{Seismic$_{1}$B$_{11}$} }
\startdata
Age (Gyr) & 4.6 & 4.6 & 4.6 & 4.735 & 4.6 & 4.6 \\
Radius ($10^{10} \, \mathrm{cm}$)  & 6.9599 & 6.95936 & 6.95866 & 6.95866 & 
6.95866 & 6.95866\\
$X_{0}$   & 0.70817 & 0.70377 & 0.70663 & 0.70958 & 0.69193 & 0.70081\\
$Z_{0}$   & 0.01959 & 0.02035 & 0.01893 & 0.01959 & 0.02035 & 0.02035\\
$(Z/X)_{0}$& 0.02766& 0.02892 & 0.02679 & 0.02761 & 0.02942 & 0.02904\\
$\alpha$  & 1.755   & 1.934   & 1.751   & 1.776   & 1.753 & 1.770\\
$(Z/X)_{s}$ & 0.0255& 0.02628 & 0.02449 & 0.02521 & 0.02684 & 0.02654\\
$Y_{s}$   & 0.2508  & 0.2508  & 0.2507  & 0.2470  & 0.2631 & 0.2549\\
$\mathrm{^{7}Li}$ (dex)& 1.14 & 1.10   & NR\tablenotemark{a} & 
NR\tablenotemark{a} & NR\tablenotemark{a} & NR\tablenotemark{a}\\
d ($R_{\odot}$)& 0.05&0.025 & 0.025& 0.025    & 0.025 & 0.025\\
N ($\mu$Hz)& 25   & 105      & 45 & 45      & 45 & 45\\
$\Omega_{0}$ (nHz)& 415 & 430 & 430 & 430 & 430 & 430\\
BCZ ($R_{\odot}$)& 0.7142 & 0.7115 &  0.7144 & 0.7121 & 0.7137 & 0.7122\\
$\mathrm{^{71}Ga}$ (SNU) & 127.1\tablenotemark{b} & 128.0 & 125.9 & 127.9& 135.1 
& 129.8\\
$\mathrm{^{37}Cl}$ (SNU) & 7.04\tablenotemark{b} & 7.47 & 7.11 & 7.47& 8.88 & 7.81\\
$\mathrm{^{8}B}$ ($10^{6} \, \mathrm{cm^{-2} \, s^{-1}}$) & 
4.98\tablenotemark{b} & 4.98 & 4.71 & 4.98 & 6.04 & 5.23\\
Atm\tablenotemark{c} & H & K & H & H & H & H\\
\enddata
\tablenotetext{a}{This model was run with a time step larger than the one 
required to derive a realistic $\mathrm{^{7}Li}$ abundance. Therefore, this 
quantity is non relevant here.}
\tablenotetext{b}{The neutrino fluxes for this model were derived with an old 
value for the capture rate of $\mathrm{^{8}B}$ neutrino by Cl ($1.06 \times 
10^{-42} \, \mathrm{cm^{2}}$ instead of $1.14 \times 10^{-42} \, 
\mathrm{cm^{2}}$), and for the $S_{17}(0)$ factor ($19$ eVb instead of $18.3$ 
eVb). The impact of updating the $S_{17}(0)$ factor is to reduce the 
$\mathrm{^{8}B}$ neutrino fluxes by about $3.5\%$.} 
\tablenotetext{c}{The atmosphere model: H for Hopf, K for Kurucz5777}
\end{deluxetable}

\subsection{Constraints Suggested by the Seismic Models}

In the previous sections we emphasize the sensitivity of $c_{s}$ to a large set 
of parameters used to work out different seismic models: that makes it difficult 
to draw firm conclusions concerning the changes in just a few parameters, since 
these modifications cannot be proved to be the unique solution. However, under 
the assumption that the major uncertainties in the solar models (for the core 
and radiative zone) are due to some reaction rates, the metallicity and the 
opacities, our seismic models shed new lights on these physical quantities. 

First, the different seismic models we computed all favor a slight increase in 
the $S(0)_{p-p}$ factor: from $0.75\%$ to $1.3\%$ depending on the model. The 
strong influence of the $p-p$ reaction was known for a long time (Turck-Chi\`eze 
\& Lopes 1993) and a raise was also recently proposed by Antia \& Chitre (1998). 
However, the increase we work out is less than theirs (due to the data improvement) and could be even less if the solar age turned out to be slightly 
larger than $4.6$ Gyr (on the contrary, a lower age implies a larger increase of 
$S(0)_{p-p}$). Incidentally, this result confirms that the calculations of 
$S(0)_{p-p}$ are quite good despite their purely theoretical base.

\begin{figure*}[htb]
\epsscale{1.0}
\plotone{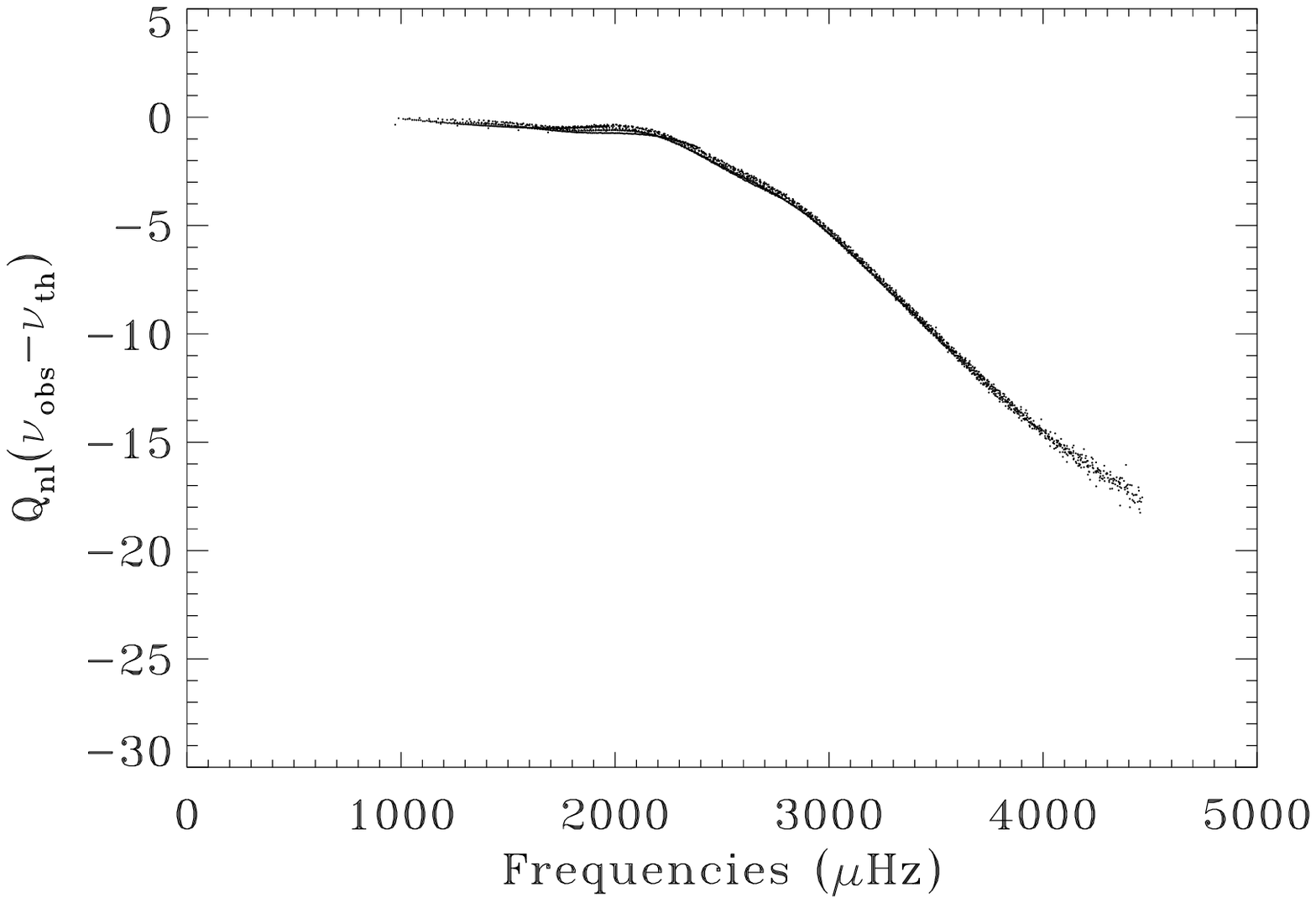}
\caption{\label{freq} {\it Weighted difference between the p mode frequencies of the 
seismic$_{1}$ model and the observed frequencies. These latter are the one of 
GOLF for $\ell \le 3$ (Garc\'\i a et al. 2001 and Bertello et al. 2000a), and 
the one of MDI for $\ell > 3$ (Rhodes et al. 1997).} }
\end{figure*}

\begin{figure*}[htb]
\epsscale{0.8}
\plotone{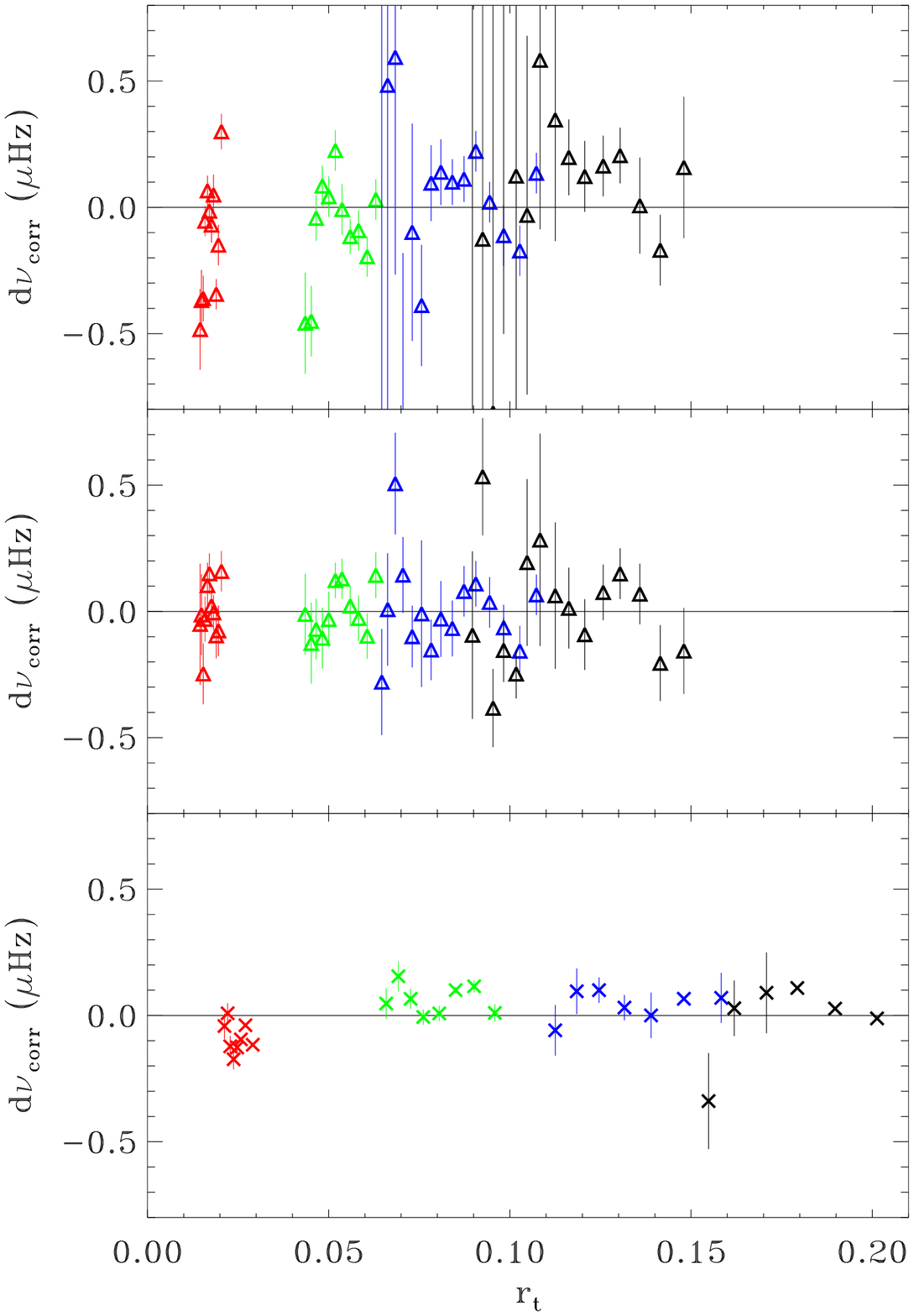}
\caption{\label{freqcomp} {\it Difference between the p mode frequencies of the 
seismic$_{1}$ model and the observed frequencies from GOLF, as a function of the internal turning point. In the upper figure: we have used the frequencies of Lazrek et al. (1997). In the middle and lower figures: Garc\'\i a et al. (2001) and Bertello et al. (2000a). For the upper and middle figures we show the modes with $n \ge 15$, for the lower figure we show $n \le 14$. We have corrected the frequency differences by the usual $Q_{nl}$ factors, and by removing the surface effects (we fitted the slope by a polynomial). Color code: the red symbols are for $\ell=0$, the green ones for $\ell=1$, the blue for $\ell=2$, and the black for $\ell=3$.} }
\end{figure*}

Concerning $Z_{0}$, an increase by $3.9\%$ is strongly favored by the 
seismic$_{1}$ model, while the seismic$_{2}$ model shows that an increase of 
$\kappa_{ross}$ combined with a decrease of $Z_{0}$ yields the same result in 
terms of $c_{s}$. Since either solutions are acceptable, we cannot favor the 
increase or decrease of $Z_{0}$. Opacities and heavy element abundances are 
closely related, and part of the error bars on $\kappa_{ross}$ are due to the 
uncertainty on $Z$.

However, this change by almost $4\%$ in $Z$ is far from trivial if you consider that the microscopic diffusion changed $Z$ by only $10\%$ since the beginning of the hydrogen burning.
Actually, the need to increase $Z_{0}$  ---in the seismic$_{1}$ model--- could be the manifestation of some forgotten hydrodynamic phenomenon. The use of both the density and rotation profiles in the core may help to solve this point. Gravity modes may also be helpful. For now, we only have a static view of the radiative region (contrary to the convective one). For instance, we cannot introduce the history of the angular momentum in the stellar equations: therefore, it seems that only hydrodynamical simulations will be able to reproduce the real Sun.
Nevertheless, the present analysis tends toward a reduced effect on the neutrino prediction when we do not correctly simulate the dynamical phenomena. Thus, this static view of the Sun remains an astrophysical problem that could have a larger impact on other stars.

Concerning $t_{\odot}$, the sound speed and density profiles change in an 
opposite way: while a raise in $t_{\odot}$ reduces the discrepancy on $c_{s}$ 
below  $0.2 R_{\odot}$, it increases it for $\rho$. If we refer to the density 
profile, which is more sensitive, it seems that $t_{\odot}=4.6$ Gyr is a 
satisfactory value for the solar age (the value needed to best reduce the 
discrepancy on $\rho$ below $0.6 R_{\odot}$ is $4.55$ Gyr). Moreover, an 
augmentation of $t_{\odot}$ is strongly disfavored by the density, while the 
sound speed does not rule it out (but does not strongly support it too).

Finally, it seems that calibrating the solar models to a radius slightly smaller 
than the usual one (by 50-125 km, this value depends on the atmosphere model) 
improves the agreement between the Sun and the models above $0.9 \, R_{\odot}$.

\section{THE OSCILLATION FREQUENCIES}

An important result of the solar models is the computation of
the oscillation frequencies for the p and g modes. We use the 
seismic$_{1}$ model: it was derived using the MHD EOS 
(Mihalas, D\"{a}ppen \& Hummer 1988) and the k5777 $T(\tau)$ law 
(derived from a Kurucz's model) for the atmosphere model (with a reconnection 
at $\tau$ = $20$ following Morel et \mbox{al.} (1994)). This is an improvement in comparison with
our previous Btz model using the 
$T(\tau)$ Hopf's law, with the CEFF EOS (Christensen-Dalsgaard \& 
D\"{a}ppen 1992). 

The Figure \ref{freq} shows the difference ---scaled by the usual $Q_{n,l}$ factors--- between the theoretical and the 
observed p-mode frequencies, up to $\ell=200$, versus the inner turning point $r_{t}$ expressed as:

\begin{equation}
\frac{w^{2}+4 \pi G \rho}{c_{s}^{2}}=\frac{L^{2}}{r_{t}^{2}}
\end{equation}

$w$ is the mode frequency, $G$ the gravity constant and $L=(\ell+0.5)$ with $\ell$ the mode degree.
We have taken the self-gravity of plane acoustic waves into account.

As usual, there is 
a very good agreement for frequencies smaller than 2 mHz and a worse one
at larger frequencies, mainly due to the difficulty to model the turbulent surface. 
The Table \ref{gmod} gives the frequencies of the low degree g modes. 

The Figure \ref{freqcomp} details the quality of the seismic data we use by showing 
the weighted difference between calculated and observed frequencies, as a function of the internal turning point. The surface effect has been removed by a polynomial fit of the general trend above 2.2 mHz.
We separate the modes in two ranges: in the first two figures we draw modes with $n > 14$ 
and in the third one, those with $n \le 14$. It appears clearly that 
the use of the higher frequency domain is not disfavored by the general trend previously
 observed. The real difficulty comes from the stochastic excitation and the mode correlated lifetime.  
 This leads to a bad determination of the frequencies 
after 8 months of measurement (\mbox{Fig.} \ref{freqcomp}a). The situation is improved by a longer 
integration time (here 1290 days)
(\mbox{Fig.} \ref{freqcomp}b) but it is obvious that the comparison computed/observed frequencies is 
better when we access to the low frequency range ($n < 15$ or $\nu < 2.2$ mHz). In this case, 
the modes have a longer lifetime and an insight into the very central core 
(which is important for high energy neutrino fluxes) could be obtained with a large increase 
of the sensitivity. Therefore, the inversion of the sound speed is better determined (even 
if the use of the sole low order modes reduce the radial accuracy), as for the density profile.
This is natural since we also reach the modes that have a mixed gravity and acoustic character.

\begin{deluxetable}{llcllcllcllcllc}
\tablewidth{0pt}
\tablecaption{G-mode Frequencies with the Seismic$_{1}$ Model \label{gmod}}
\tablehead{\colhead{$\ell$} & \colhead{$n$} & \colhead{$\nu$ ($\mu$Hz)} & 
\colhead{$\ell$} & \colhead{$n$} & \colhead{$\nu$ ($\mu$Hz)} & \colhead{$\ell$} & \colhead{$n$} & \colhead{$\nu$ ($\mu$Hz)} & \colhead{$\ell$} & \colhead{$n$} & \colhead{$\nu$ ($\mu$Hz)} & \colhead{$\ell$} & \colhead{$n$} & \colhead{$\nu$ ($\mu$Hz)}}
\startdata
1   &    -6   &    95.47  & 3   &   -16   &    93.12  & 4  &    -13  &     136.49 & 5  &    -14  &     150.83 & 6   &   -17  &     147.70\\
1   &    -5   &    109.28 & 3   &   -14   &    104.21 & 4  &    -12  &     145.31 & 5  &    -13  &     159.55 & 6   &   -15  &     162.42\\
1   &    -4   &    127.79 & 3   &   -13   &    110.84 & 4  &    -11  &     155.27 & 5  &    -12  &     169.27 & 6   &   -14  &     170.87\\
1   &    -3   &    153.25 & 3   &   -12   &    118.37 & 4  &    -10  &     166.60 & 5  &    -11  &     180.14 & 6   &   -13  &     180.19\\
1   &    -2   &    191.55 & 3   &   -11   &    126.96 & 4  &     -9  &     179.62 & 5  &    -10  &     192.39 & 6   &   -12  &     190.49\\
1   &    -1   &    262.73 & 3   &   -10   &    136.85 & 4  &     -8  &     194.59 & 5  &     -9  &     206.31 & 6   &   -11  &     201.94\\
2   &   -11   &    94.91  & 3   &    -9   &    148.33 & 4  &     -7  &     211.92 & 5  &     -8  &     222.13 & 6   &   -10  &     214.69\\
2   &   -10   &    102.68 & 3   &    -8   &    161.72 & 4  &     -6  &     231.62 & 5  &     -7  &     240.19 & 6   &    -9  &     229.05\\
2   &    -9   &    111.82 & 3   &    -7   &    177.46 & 4  &     -5  &     250.39 & 5  &     -6  &     260.12 & 6   &    -8  &     245.20\\
2   &    -8   &    122.63 & 3   &    -6   &    195.93 & 4  &     -4  &     265.08 & 5  &     -5  &     272.04 & 6   &    -7  &     263.41\\
2   &    -7   &    135.59 & 3   &    -5   &    217.07 & 4  &     -3  &     291.42 & 5  &     -4  &     288.27 & 6   &    -6  &     283.04\\
2   &    -6   &    151.26 & 3   &    -4   &    238.35 & 4  &     -2  &     328.10 & 5  &     -3  &     316.38 & 6   &    -5  &     289.06\\
2   &    -4   &    194.06 & 3   &    -3   &    261.31 & 4  &     -1  &     368.29 & 5  &     -2  &     350.90 & 6   &    -4  &     308.82\\
2   &    -3   &    222.02 & 3   &    -2   &    296.50 & 5  &    -17  &     129.44 & 5  &     -1  &     385.55 & 6   &    -3  &     335.81\\
2   &    -2   &    256.09 & 3   &    -1   &    340.07 & 5  &    -16  &     135.89 & 6  &    -19  &     135.32 & 6   &    -2  &     367.39\\
2   &    -1   &    296.38 & 4   &    -14  &    128.64 & 5  &    -15  &     142.98 & 6  &    -18  &     141.25 & 6   &    -1  &     396.69\\  
\enddata
\end{deluxetable}

\begin{figure*}[htb]
\epsscale{0.8}
\plotone{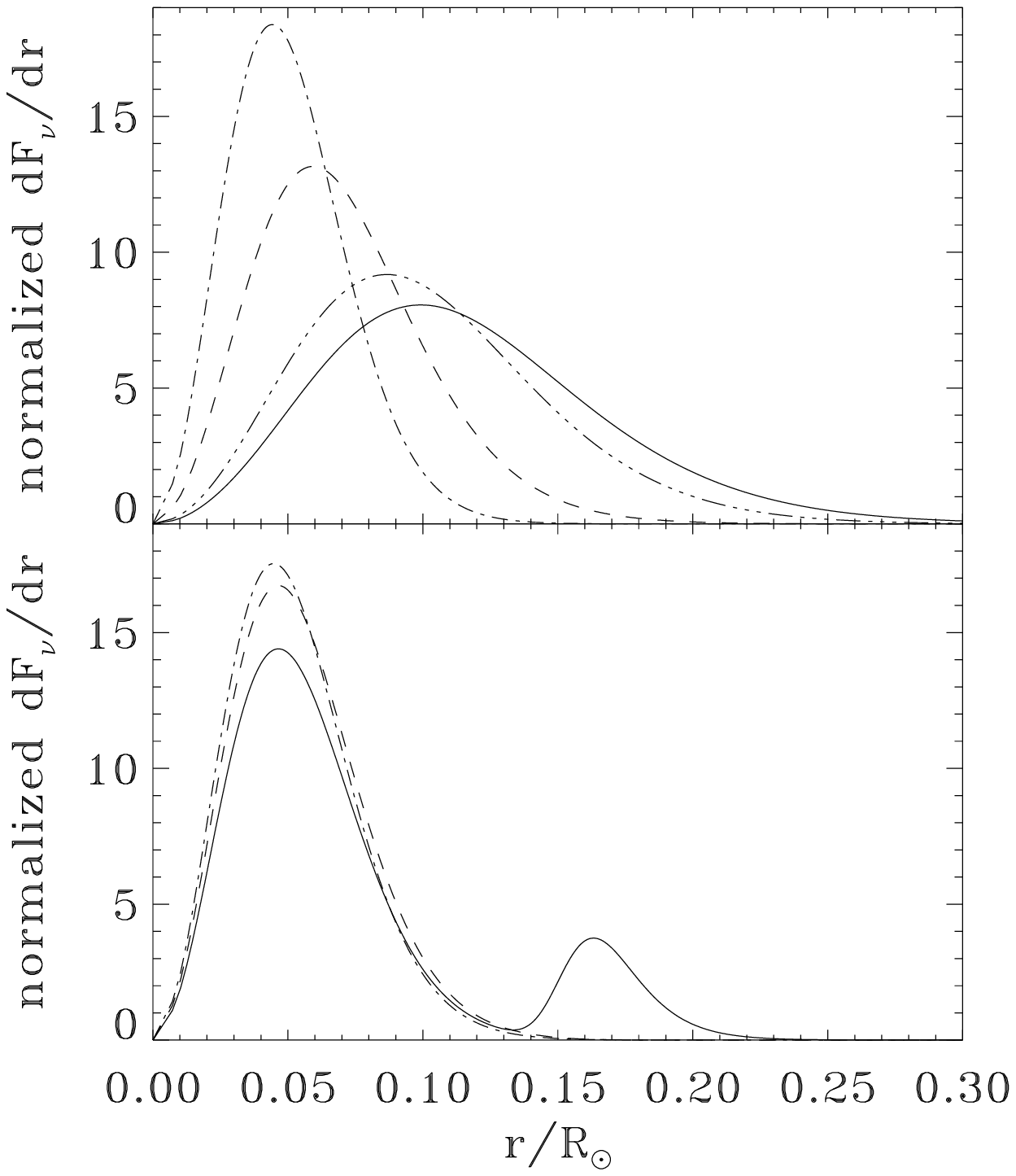}
\caption{\label{fluxesn} {\it Neutrino production as a function of the 
fractional radius.
In the upper figure, we have drawn the p-p (plain curve), $\mathrm{^{8}B}$ 
(dot-dashed curve), $\mathrm{^{7}Be}$ (dashed curve), and the pep 
(dot-dot-dot-dashed curve) neutrinos.
In the lower figure, the $\mathrm{^{13}N}$ (plain curve), $\mathrm{^{15}O}$ 
(dashed curve), and $\mathrm{^{17}F}$ (dot-dashed curve) neutrino production is 
shown.
For each neutrino ``type'', we have drawn $(1/F_{t}) \, (dF/dr)$ where $F$ is 
the flux in $\mathrm{s^{-1}}$, $r$ the fractional radius, and $F_{t}$ the total 
flux for this neutrino type (integrated over the entire Sun). } }
\end{figure*}

\section{SEISMIC PREDICTED NEUTRINO FLUXES}

\subsection{The Solar Neutrino Predictions}

Since the neutrino production is not pointlike, the precise calculation of the 
expected emitted $\nu_{e}$ flux  requires to know where the neutrinos are produced. 
The Figure \ref{fluxesn} recalls the production zones for the p-p, pep, $\mathrm{^{8}B}$, 
$\mathrm{^{7}Be}$, $\mathrm{^{13}N}$, $\mathrm{^{15}O}$ and $\mathrm{^{17}F}$ 
neutrino fluxes. As it clearly appears on the figure, most of them depend on the very central part. Now we recall that 
the first radius we have for the sound speed is at $0.07 \pm 3.6\% R_{\odot}$ (Table 1 of Turck-Chi\`eze et al. 2001b), and
 only gravity modes may improve this situation.

Neutrino predicted fluxes have been calculated for the three seismic models (see Table 1). 
We may note that they do not differ significantly.  Moreover, they are only 
slightly different from our previous predictions in Brun et \mbox{al.} (1999) or from the recent solar models (\mbox{e.g.} Bahcall et \mbox{al.} 2001). 
This is logical since the last modifications we introduced  are minor. 
In Turck-Chi\`eze (2001), we recall the progress made to stabilize these fluxes.  
Detailed neutrino capture predictions are shown for the seismic$_{1}$ model (see Table \ref{tab2}). 

The present predictions are of great interest (in comparison with those 
obtained ten years ago) since they include 
helioseismic data that validate the updated physics at the needed level 
(see Table 2 of Turck-Chi\`eze et al. 2001a). 
The emitted neutrino 
fluxes we predict are not only theoretical but also partly deduced from precise 
 seismic ``observations'' of the solar core, and they confirm the discrepancy 
 between predictions and detections, favouring the neutrino flavor transformation. 
 The next step is to sum the fluxes of these different flavors received on Earth.
Such data are now available thanks to the SNO experiment.

\begin{deluxetable}{lccccccc}
\tabletypesize{\footnotesize}
\tablewidth{0pt}
\tablecaption{Neutrino Fluxes of the Seismic$_{1}$ Model \label{tab2}}
\tablehead{\colhead{} & \colhead{pp} & \colhead{pep} & 
\colhead{$\mathrm{^{7}Be}$} & \colhead{$\mathrm{^{8}B}$} & 
\colhead{$\mathrm{^{13}N}$} & \colhead{$\mathrm{^{15}O}$} & 
\colhead{$\mathrm{^{17}F}$} }
\startdata
Sun\tablenotemark{a} & 1.673$\times \, 10^{38}$ & 3.936$\times \, 10^{35}$ & 
1.372$\times \, 10^{37}$ & 1.408$\times \, 10^{34}$ & 1.631$\times \, 10^{36}$ & 
1.404$\times \, 10^{36}$ & 8.718$\times \, 10^{33}$ \\
Earth \tablenotemark{b}& 5.916$\times \, 10^{10}$ & 1.392$\times \, 10^{8}$ & 
4.853$\times \, 10^{9}$ & 4.979$\times \, 10^{6}$ & 5.767$\times \, 10^{8}$ & 
4.967$\times \, 10^{8}$ & 3.083$\times \, 10^{6}$ \\
$\mathrm{^{71}Ga}$\tablenotemark{c} & 69.34 & 2.840 & 34.79 & 11.95 & 3.483 & 
5.648 & 3.512$\times \, 10^{-2}$ \\
$\mathrm{^{37}Cl}$\tablenotemark{c} & 0.000 & 2.228$\times \, 10^{-1}$ & 1.155 & 
5.676 & 9.573$\times \, 10^{-2}$ & 3.283$\times \, 10^{-1}$ & 2.057$\times \, 
10^{-3}$ \\
\cutinhead{Capture Predictions for \tablenotemark{d}}
\sidehead{$\mathrm{^{71}Ga}$ : $128.1 \pm 8.9$ SNU,  $\mathrm{^{37}Cl}$ : $7.48 
\pm 0.97$ SNU, water experiments : $4.98 \pm 0.73 \, \times 10^{6}$ 
$\mathrm{cm^{-2} \, s^{-1}}$}
\enddata
\tablenotetext{a,b}{flux emitted by the Sun, and received on Earth, in units of 
neutrinos $\times \, \mathrm{cm^{-2} \, s^{-1}}$}
\tablenotetext{c}{in SNU, Solar Neutrino Unit: 1 SNU $=$ $10^{-36}$ captures per 
second and per target atom}
\tablenotetext{d}{The error bars on the predictions are derived from table 2 of 
Turck-Chi\`eze et \mbox{al.} (2001b)}
\end{deluxetable}

\subsection{Comparison of the $\mathrm{^8B}$ Emitted Flux with the SNO \& Superkamiokande Results}

Ahmad et 
\mbox{al.} (2001) announced the first results of the SNO collaboration
on charged current (CC) 
and elastic scattering off electrons (ES) reactions. 

The CC reaction is only sensitive to electron-type neutrinos, while the ES reaction, 
also measured in the Superkamiokande experiment, is sensitive to all flavors. The 
measured $\mathrm{^{8}B}$ neutrino fluxes for the CC and ES reactions are:
$$\Phi^{CC}_{SNO}(\nu_{e})= 1.75\pm0.07({stat.})^{+0.12}_{-0.11}({sys.}) 10^{6} \mathrm{cm^{-2}s^{-1}}$$
$$\Phi^{ES}_{SNO}(\nu_{x}) =  2.39\pm0.34({stat.})^{+0.16}_{-0.14}({sys.}) 10^{6} \mathrm{cm^{-2}s^{-1}}$$

The present statistical error on the elastic scattering is still large and the 
 sensitivity to the $\nu_{\mu}$, $\nu_{\tau}$ or anti $\nu_{\mu}$, $\nu_{\tau}$ (only $16\%$ 
in comparison to $100\%$ for $\nu_e$) is low. Therefore, it is not possible yet 
to properly extract the different flavors from the SNO experiment alone, despite the difference in the two measured fluxes.
Fortunately, the ES result is in agreement with the measure from the SK experiment (Fukuda et \mbox{al.} 2001):
$$\Phi^{ES}_{SK}(\nu_{x}) =  2.32\pm0.03({stat.})
^{+0.08}_{-0.07}({sys.}) 10^{6} \mathrm{cm^{-2}s^{-1}}$$

This flux is 
equal to $\simeq 47-48\%$ of the seismic$_{1}$ model $\mathrm{^{8}B}$ neutrino 
prediction (Turck-Chi\`eze et al. 2001b):
$$ \Phi(\nu_{x})=4.98 \pm 0.73 \times 10^{6} \mathrm{cm^{-2}s^{-1}}$$
It is widely known that the $\mathrm{^{8}B}$ neutrino flux is the most difficult to predict
and is very dependent on the physics of the core. A fundamental improvement  achieved by 
the seismic model is the 
reduction of the error bars in the neutrino prediction, through the reduction of 
the uncertainty on the $p-p$ reaction rate, the $(Z/X)_{s}$ ratio... This point 
has already been discussed in the Table 2 of Turck-Chi\`eze et \mbox{al.} (2001b) 
that gives the detailed uncertainties on neutrino predictions. We point out that 
the neutrino fluxes derived here are not only 
``observational''  (to a certain extent) but are also affected by reduced 
uncertainties. They also include the recent rejection of several astrophysical 
solutions to the neutrino puzzle proposed in the past.

Nevertheless, the $S(0)_{17}$ factor has no impact 
on the structure and was  badly determined more than 4 years ago (see Turck-Chi\`eze 2001).
For instance, Adelberger et \mbox{al.} (1998) propose 
$S(0)_{17}=19^{+4}_{-2}$ eV b. 
But the experimental measurements have been largely improved. In our models, we use 
the value of Hammache et al. (1998) of $18.5 \pm 1$ eV b. Unfortunately, the recent result of 
Junghans et al. (2002)
of $22.3 \pm 0.7 $ eV b, is only marginally in agreement with the previous one. Thus the uncertainty on the present predicted neutrino flux, including the one of the seismic$_{1}$ model, 
could be a bit larger today.

Actually, the major result of the last year is the one obtained by the SNO collaboration when they add the
estimated number of detected muon and tau type neutrinos to the electronic ones (using the SK results); they find:

$$\Phi(\nu_{x})=5.44\pm0.99 \times 10^{6} \mathrm{cm^{-2}s^{-1}}$$
a result very consistent with the previous value validated by the present helioseismic data.

This strongly favors or even ``proves'' the existence of neutrino oscillations: a part of the $\nu_{e}$ must be converted 
into $\nu_{\mu}$ and $\nu_{\tau}$, as far as new seismic results do not lead to other conclusion.
Actually the present study stays in the framework of a static solar core, with classical phenomena.
This representation is compatible with the present seismic results but we still need to extract the rotation
and magnetic field in the radiative zone to check this assumption.

\subsection{The Neutrino Oscillation Related Quantities}

\begin{figure*}[htb]
\epsscale{0.8}
\plotone{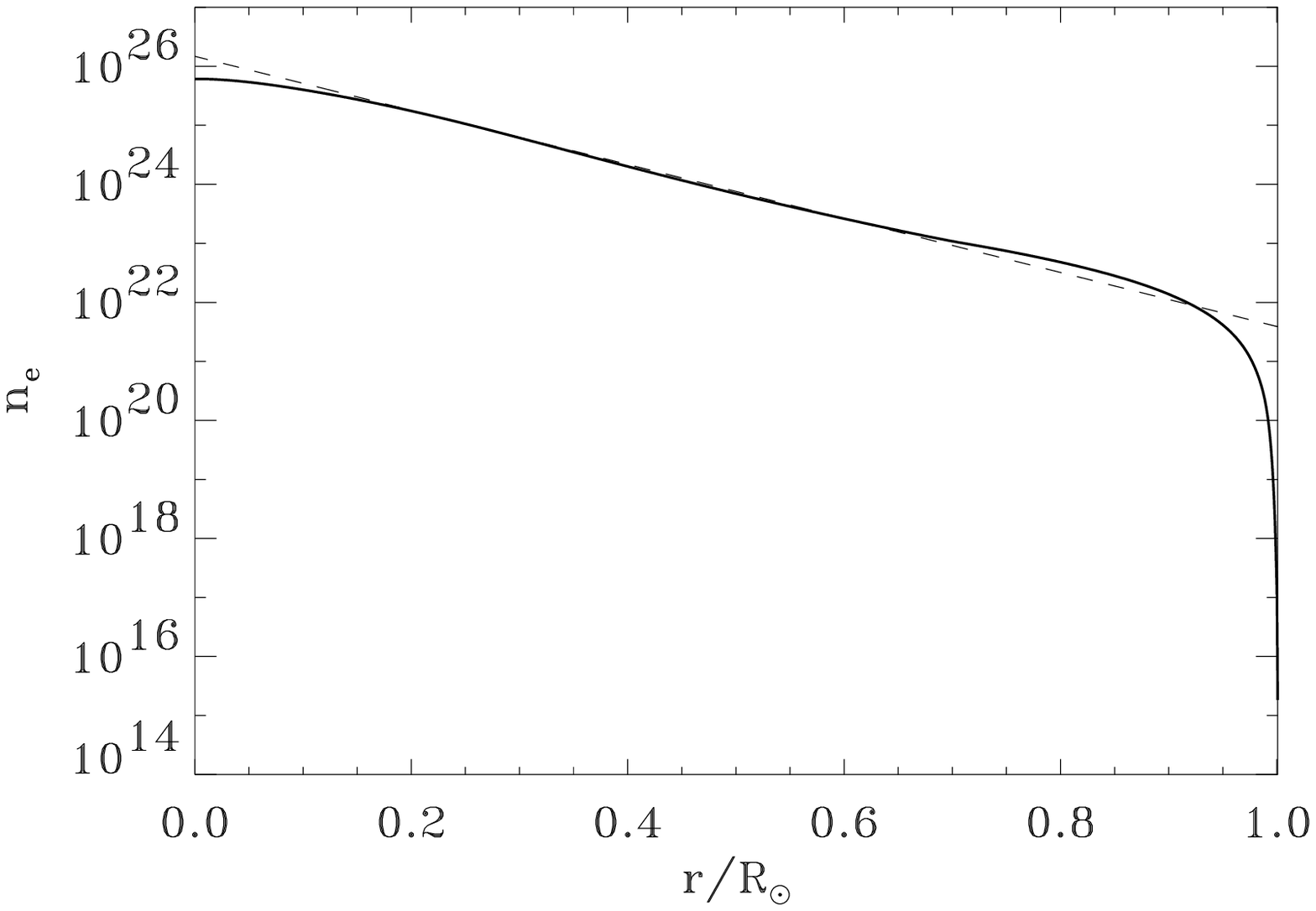}
\plotone{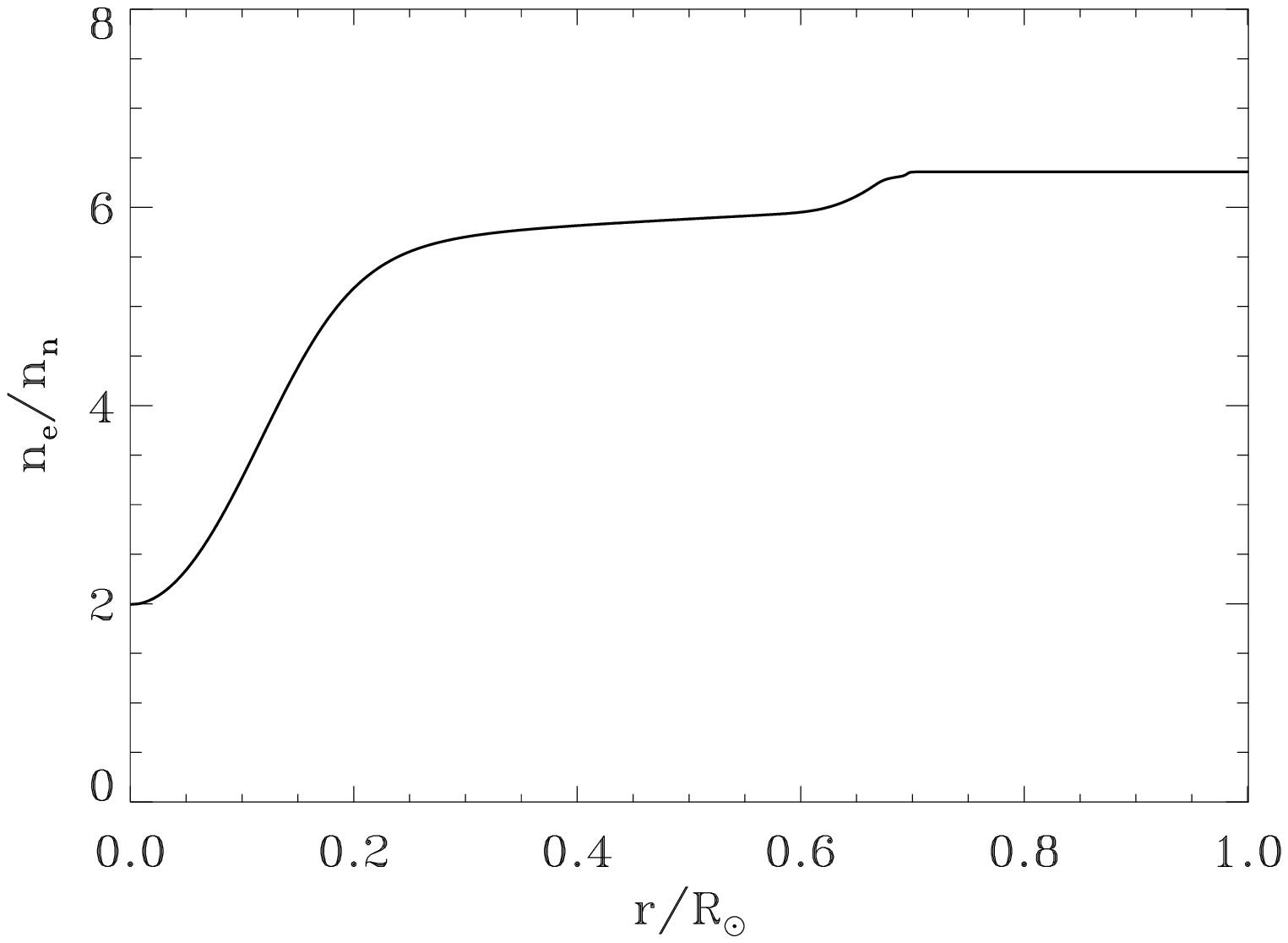}
\caption{\label{nenn} {\it Upper figure: the electron number density for the seismic$_{1}$
model. The approximation proposed by Bahcall (for instance Bahcall et \mbox{al.} 
2001) has also been drawn (dashed curve). Lower figure: the neutron number density for the seismic$_{1}$ model 
(we show the $n_{e}/n_{n}$ ratio). } }
\end{figure*}

\subsubsection{The Electron Number Density}

Following this framework, a part of the neutrino oscillations could be explained 
by the well-known Mikheyev-Smirnov-Wolfenstein effect (MSW effect, see 
\mbox{e.g.} Mikheyev \& Smirnov 1986): an electron type neutrino may undergo a 
resonant oscillation in the Sun and then be converted into a muon or tau type 
neutrino. This effect assumes that the neutrinos 
have masses and that the flavor eigenstates are different from the mass 
eigenstates.

For the sake of simplicity, we suppose two flavors ($\nu_{e}$ and $\nu_{\mu}$) and two mass eigenstates ($\nu_{1}$ and $\nu_{2}$). We assume that $\nu_{1}$ is close to $\nu_{e}$, and $\nu_{2}$ is close to $\nu_{\mu}$. When an electron type neutrino is created in the solar core, the flavor eigenstates are almost 
``propagation eigenstates'', because of the high electron number density ($n_{e}$). But during the propagation from the neutrino toward 
the solar surface, $n_{e}$ decreases and the flavor states are no more 
propagation eigenstates: the neutrino state vector starts to oscillate around 
the new propagation eigenstate. This latter changes with the further decrease of 
$n_{e}$, and gets closer to $\nu_{2}$ (initially, it was close to $\nu_{e}$). 
Depending on the variation of $n_{e}$, this change is more or less rapid and 
said adiabatic or not. In the case of an adiabatic change, the probability for an emitted electron type neutrino
 to be detected as a muon type 
neutrino at the solar surface is rather simple to calculate. In the non-adiabatic case, the conversion probability is more tricky to work out.
In both cases, the electron number density is needed with a high accuracy all 
along the solar radius, to compute the conversion probabilities.
 
Therefore, we derive $n_{e}$ for the seismic$_{1}$ model. To do so, we use the precise mass fractions returned by CESAM: we know the mass fractions for the $\mathrm{^1H}$, $\mathrm{^2H}$, $\mathrm{^3He}$, $\mathrm{^4He}$, $\mathrm{^7Li}$, $\mathrm{^7Be}$, $\mathrm{^9Be}$, $\mathrm{^{12}C}$, $\mathrm{^{13}C}$, $\mathrm{^{14}N}$, $\mathrm{^{15}N}$, $\mathrm{^{16}O}$, \& $\mathrm{^{17}O}$ chemical elements, plus an extra element (with $A=28$ and $Z=13$), as a function of the fractional radius. We just have to derive the number fractions of all these elements, and multiply these number fractions by the electron number of the corresponding chemical element. By adding all the terms we obtained this way, we determine the electon number density as a function of the radius. 
The result is drawn in the upper part of \mbox{Fig.} \ref{nenn}.

\subsubsection{The Neutron Number Density}

If the neutrino has a magnetic moment (either a dipole or/and transition moments), 
it might interact with the solar magnetic field. Provided that the magnetic 
moment and the magnetic field are large enough, this field could flip the spin 
of the neutrino: a left-handed neutrino could become right-handed. Moreover, the possible flavor transition magnetic moments 
could result in a spin-flavor precession: the neutrino could change both its 
chirality and its flavor. This double precession could be matter-enhanced 
through the interactions of neutrino with the electrons, protons and neutrons: 
this is the Resonant Spin-Flavor Precession process (RSFP, e.g. Lim \& Marciano 
1988). To compute the conversion probabilities for the RSFP, the neutron number 
density $n_{n}$ is needed (see the lower part of Fig. \ref{nenn}): we derive it the same way as the electon number density, but instead of using the electron number of each chemical element, we use its neutron number. It is important to note that both the MSW and RSFP processes could cohabit inside the Sun.

The detailed $n_{e}$ and $n_{n}$ profiles are available with the seismic$_{1}$
model on the web site whose address was previously mentioned, in order to 
calculate quantities related to the neutrino oscillations.

\section{UNSOLVED PROBLEMS IN THE SEISMIC MODELS}

Despite the overall agreement in the sound speed between the Sun and the seismic 
models below $0.6 R_{\odot}$, two regions of our star remain poorly described: 
the tachocline region and the upper layers, both are expected to undergo dynamic 
effects. Moreover the central rotation law is not taken into account in this analysis 
(Garcia, R. A., Couvidat, S., Turck-Chi\`eze, S. et al., {\it in preparation}).

In the superadiabatic region it is well known that the turbulent 
pressure becomes non negligible compared to the gas pressure. Moreover, the 
tachocline and the upper solar layers are shear layers in which rotation rates 
change rapidly (e.g. see the rotation curve of Howe et \mbox{al.} 2000). A 1D 
stellar evolution code cannot afford an efficient treatment of these dynamic 
regions. Neither the rotation of the Sun nor its magnetic field are taken into 
account, whereas it is widely thought that the tachocline is the base of the 
magnetic dynamo process. Of course, the neutrino emission and the solar core 
physics are rather insensitive to the tachocline and beyond, but the neutrino 
behaviour may depend on these layers. Progresses are needed to connect interior 
phenomena related to these neutrinos, to what is observed out of the star. 
If the neutrinos have a magnetic moment, they could interact with the 
solar interior magnetic field.
In the next section we focus on the solar large-scale magnetic field: we add 
magnetic pressure and derive new solar models.

\subsection{The Magnetic Field Tested by the Sound Speed}

The goal of this analysis is to test the sensitivity of $c_{s}(r)$ to the magnetic field and constrain 
the amplitude of this field. We also check (and use) the sensitivity of the 
density.

Basically, the presence of a magnetic field changes the wave velocity in two 
ways:
\begin{itemize}
\item it changes the gas pressure because of the hydrostatic equilibrium;
\item it adds a part of the Alfv\'en speed $v_{a} = B^{2}/(4 \pi \rho)$ (in cgs units) to the wave velocity.
\end{itemize}

Thus, any magnetic field should be imprinted in the ``magneto-acoustic'' wave 
velocity. Of course, the 3D structure of the field is of great importance since 
the angle between the field lines and the seismic waves determines the way these 
latter are accelerated: the wave velocity is no more an isotropic quantity. 
Unfortunately, we cannot account for the field structure with a 1D stellar code, 
and so we just add a magnetic pressure term $P_{mag}$ in the stellar structure 
equations: $P_{mag}=B^{2}/(8\pi)$ (in cgs units).
Therefore, the basic hydrostatic equilibrium equation is changed 
into:

\begin{equation}
\frac{dP_{gas}}{dm} = -\frac{G M}{4\pi r^{4}} -\frac{dP_{mag}}{dm}
\end{equation}

The main problem is to choose an appropriate magnetic field $\bf{B(r)}$ for the 
solar interior, since little is known about the inner field. Roughly speaking, 
it seems that large scale magnetism may be important in three regions: the 
radiative zone, the tachocline, and just below the solar surface.

\subsection{Origin of a Large Scale Magnetic Field}

Parker presented the hydromagnetic dynamo theory for the Sun in 1955. The 
$\alpha-\omega$ kinematic dynamo needs both differential latitudinal rotation 
(parameter $\omega$) and turbulent ---cyclonic--- movements (parameter 
$\alpha$). The differential rotation produces a toroidal magnetic field from a 
poloidal one, while the cyclonic movements slow down the lift and twist of this 
toroidal field which has enough time to get amplified. The lift of the field is 
due to the magnetic buoyancy, while the twist results from the Coriolis force. 
Both these actions induce a poloidal field from the toroidal one, thus restoring 
the initial field. This basic picture has been widely reviewed as different 
dynamo processes were developped and improved. For instance, let us mention the 
Babcock-Leighton dynamo models that regenerate the poloidal field by the 
eruption of the toroidal field at the solar surface (\mbox{e.g.} see Durney 1997 
for a modified Babcock-Leighton dynamo analysis). 
It is now commonly believed that the seat of the (main) solar dynamo process is 
the tachocline. A small seed poloidal field ---a legacy of the PMS evolution--- 
in the radiative interior should have given birth to a magnetic field with 
poloidal and toroidal components. Although this scenario seems widely accepted, 
the situation is rather confused when going into details. The seat of the 
poloidal field regeneration is not located, it could be the tachocline (as is induced by the 
``Parker'' dynamo process), or just below the solar surface (as proposed 
by the Babcock-Leighton dynamo).
Moreover, it is unclear whether or not the toroidal part of the field prevails 
in the solar interior: the surface activity proves that the toroidal field is 
larger than the poloidal one in the upper layers of the Sun, but only a few 
clues of what happens in the deep interior are known. 
We just consider toroidal fields in this paper.

\subsection{Simulated Magnetic Profiles}

Following Dzitko (1995) we simulate fields as:

\begin{figure*}[pht*]
\epsscale{1.0}
\plotone{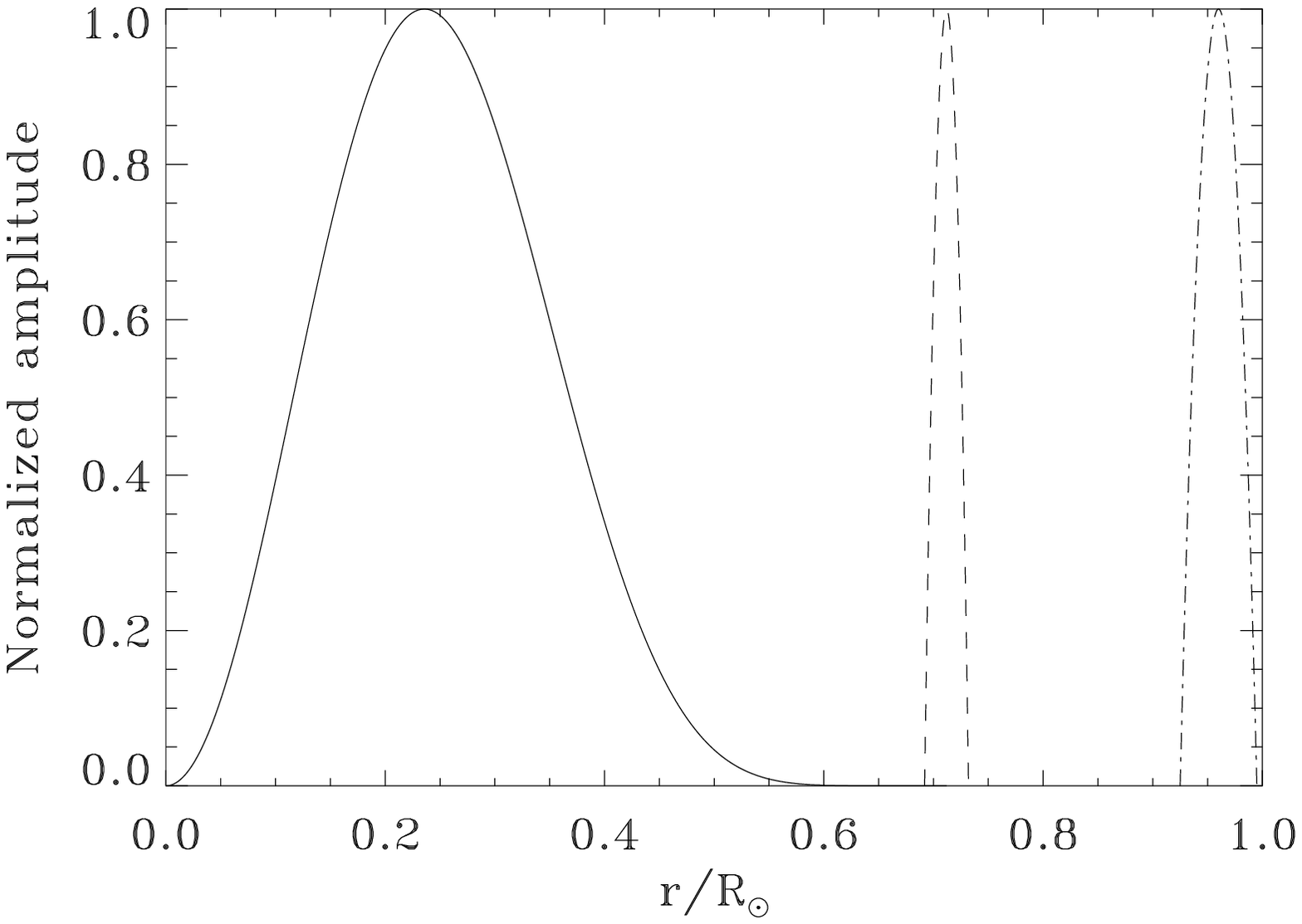}
\caption{\label{mag} {\it Magnetic fields simulated: in the radiative interior 
(plain curve), in the tachocline (dashed curve), in the upper layers (dot-dashed 
curve). The amplitudes of the fields have been normalized.}}
\end{figure*}

\begin{equation}
\mathbf{B_{\phi}} = a(r) \frac{d}{d \theta}P_{k}(cos \theta) \, 
\mathbf{e_{\phi}}
\end{equation}

With the spherical coordinates $(r,\theta,\phi)$.
$P_{k}(cos\theta)$ is a Legendre polynomial of degree $k$. 
 
For argument's sake, we assume $k=2$, meaning that the field is quadrupolar (in 
accordance with the surface-magnetism manifestations). The presence of a 
magnetic pressure modifies in a direct way the wave velocity: we must take into account 
the Alfv\'en speed. Since we are only interested in the radial velocity and a 
toroidal field is perpendicular to the radial direction, the new wave velocity 
is $\sqrt{c_{s}^{2}+v_{a}^{2}}$ (on the figures, we continue to note this 
quantity ``$c_{s}$'' for convenience, eventhough it is no more the sound speed, 
strictly speaking). 
For the function $a(r)$, two profiles are considered, following  Gough \& 
Thompson (1990).\\

First, to simulate a magnetic field in the radiative zone, we choose: 

\begin{equation}
a(r) = \left\{ \begin{array}{ll}
K_{\lambda}(\frac{r}{r_{0}})^{2}(1-(\frac{r}{r_{0}})^{2})^{\lambda} & \textrm{if 
$r \le r_{0}$} \\
0 & \textrm{otherwise}
\end{array} \right.
\end{equation}

With $K_{\lambda}=(1+\lambda) (1+1/\lambda)^{\lambda} \, B_{0}$, $r_{0}= 0.712 
R_{\odot}$ being approximately the BCZ and $\lambda = 10 r_{0} +1$. 
$B_{0}$ ---the highest intensity of the field--- is set to different values (see 
Table \ref{tab3}). The addition of the magnetic pressure is made when the Sun 
enters the ZAMS. The intensities of these fields are very high, compared to the 
prediction of Gough \& MacIntyre (1998) who claim for a (poloidal) field with an 
amplitude of $\approx 0.1$ T in the deep interior. Moreover, the virial theorem 
rules out fields larger than $\approx 10^{4}$ T (Mestel \& Weiss 1987). The 
maximum values of the $P_{mag}/P_{gas}$ ratio (hereafter $\beta^{-1}$) are 
available in Table \ref{tab3}, for all the models discussed here. We could also 
give the $v_{a}/c_{s}$ ratio, but $\beta$ and this ratio are closely 
related: $v_{a}/c_{s} = \sqrt{2 / (\beta \, \Gamma_{1})}$. Both contributions of 
the magnetic field to the change in $c_{s}$ --- the indirect change through the 
modification of the solar structure, and the direct change through the addition 
of $v_{a}$--- depend on the $\beta$ values.\\

Second, we also utilize as $a(r)$ profile:

\begin{equation}
a(r) = \left\{ \begin{array}{ll}
B_{0} (1-(\frac{r-r_{0}}{d})^{2}) & \textrm{if $\vert r-r_{0}\vert \le d$} \\
0 & \textrm{otherwise}
\end{array} \right.
\end{equation}

\begin{deluxetable}{lccc}
\tablecaption{Models with Magnetic Field \label{tab3}}
\tablehead{\colhead{Name} & \colhead{$B_{0}$ (T)} & 
\colhead{Center\tablenotemark{a} ($R_{\odot}$)} & 
\colhead{$(P_{mag}/P_{gas})_{max}$}} 
\tablewidth{0pt}
\startdata
seismic$_{1}$B$_{1}$ & $10^{4}$  & $0.236$ & $2.85 \times 10^{-2}$ \\
seismic$_{1}$B$_{11}$& $5 \times 10^{3}$  & $0.236$ & $6.96 \times 10^{-3}$ \\
seismic$_{1}$B$_{12}$& $3 \times 10^{3}$  & $0.236$ & $2.49 \times 10^{-3}$ \\
seismic$_{1}$B$_{13}$& $1 \times 10^{3}$  & $0.236$ & $2.80 \times 10^{-4}$ \\
seismic$_{1}$B$_{2}$ & $30$  & $0.712$ & $6.15 \times 10^{-5}$ \\
seismic$_{1}$B$_{21}$& $50$  & $0.712$ & $1.71 \times 10^{-4}$ \\
seismic$_{1}$B$_{3}$ & $2$   & $0.96$ & $1.34 \times 10^{-4}$\\
seismic$_{1}$B$_{31}$& $3$   & $0.96$ & $3.02 \times 10^{-4}$\\
\enddata
\tablenotetext{a}{radius at which $P_{mag}$ is maximum}
\end{deluxetable}

With $d$ the half-width of the zone with a magnetic field, and $r_{0}$ its 
center. To simulate a magnetic field in the tachocline, parameters are set to: 
$d = 0.02 R_{\odot}$ and the radius $r_{0}$ of transition between radiative and 
convective zones varies along the solar evolution. We set $B_{0}$ to $30$ T  
according to Antia et \mbox{al.} (2000) for the seismic$_{1}$B$_{2}$ model, and 
$50$ T for the seismic$_{1}$B$_{21}$ model.

Finally, this profile is also used to simulate a possible field in the upper 
solar layers. Such a field was hinted by Antia et \mbox{al.} (2000) from an 
analysis of the Global Oscillation Network Group (GONG) and MDI data. To test a 
field anchored at $0.96 R_{\odot}$, parameters are set to $r_{0} = 0.96 
R_{\odot}$ and $d = 0.035 R_{\odot}$. $B_{0}$ is set to $2$ (the value proposed by Antia et \mbox{al.} 2000) and $3$ T.
The fields at the BCZ and in the upper layers are both added at $85$ million 
years, when the solar core is no more convective.

\begin{figure*}[htb]
\epsscale{0.8}
\plotone{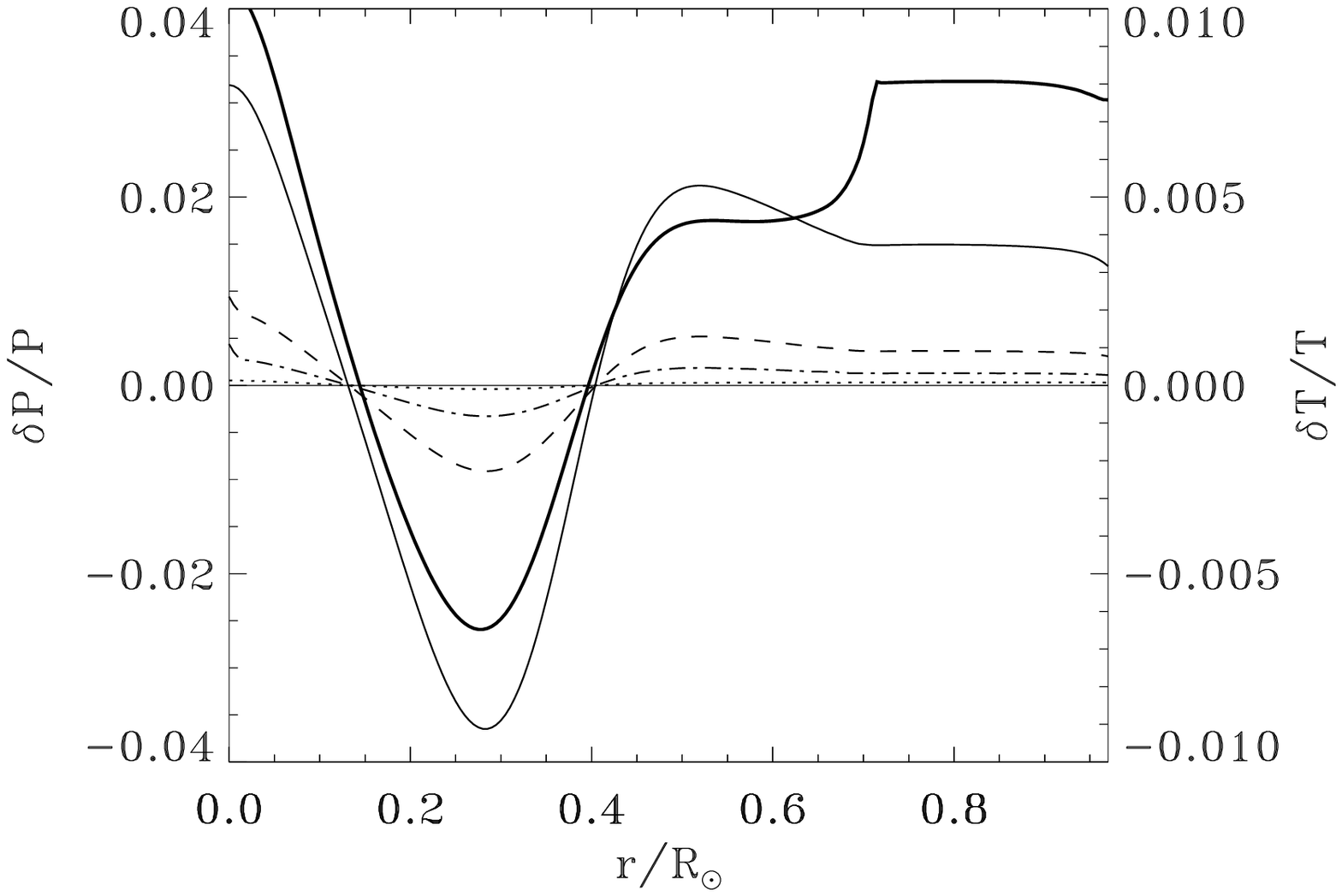}
\plotone{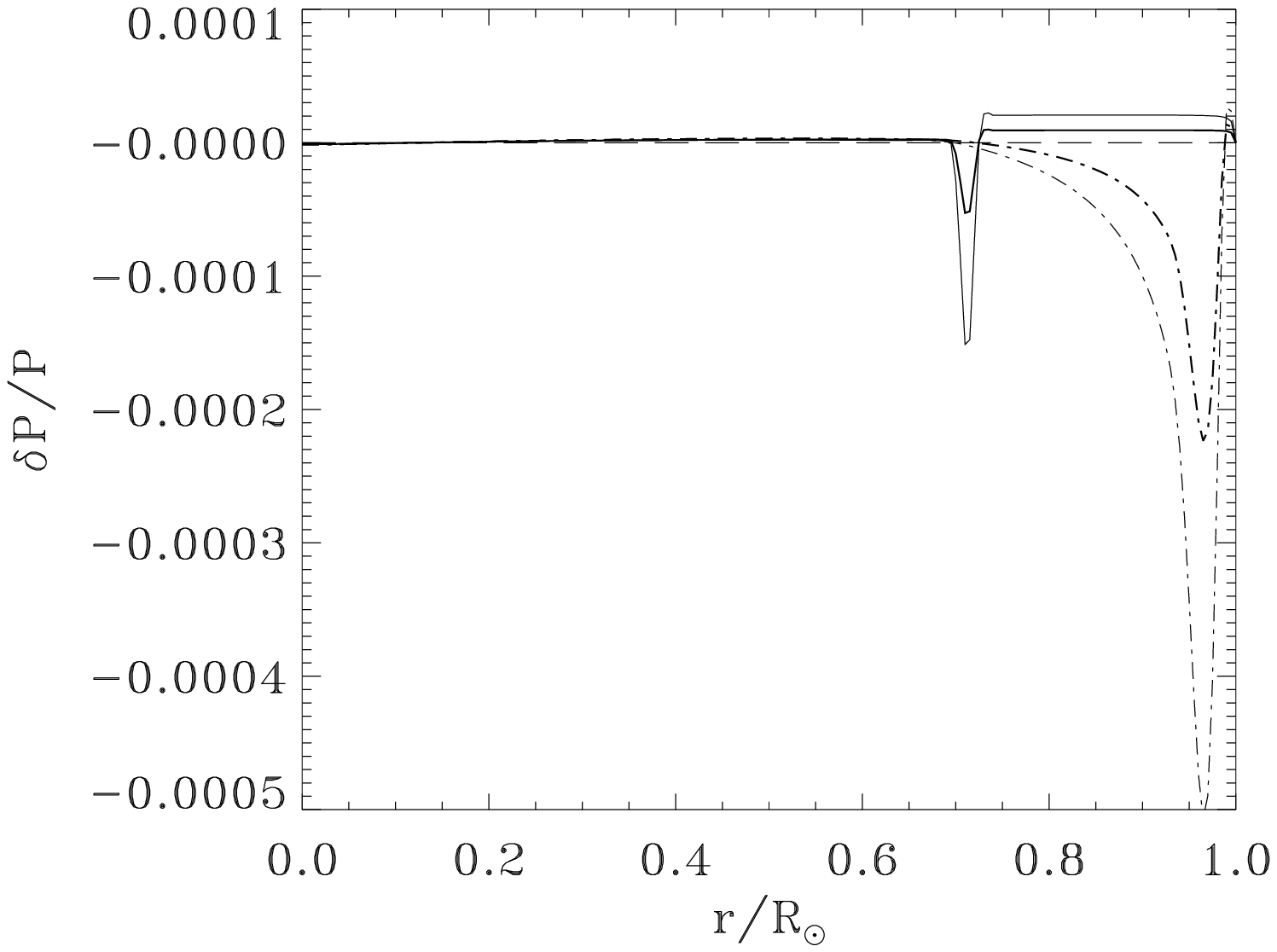}
\caption{\label{pressionB1} {\it Upper figure: difference in the gas pressure between the 
seismic$_{1}$B$_{1}$ and the seismic$_{1}$ models (thin plain curve). Same 
figure for the seismic$_{1}$B$_{11}$ (dashed curve), the seismic$_{1}$B$_{12}$ 
(dot-dashed curve), and the seismic$_{1}$B$_{13}$ (dotted curve) models.
Superimposed is the difference in the temperature between the 
seismic$_{1}$B$_{1}$ and the seismic$_{1}$ models (thick plain curve).
Lower figure: Same 
figure for the seismic$_{1}$B$_{2}$ model (thick plain curve), the seismic$_{1}$B$_{21}$ (thin plain curve), the 
seismic$_{1}$B$_{3}$ (thick dot-dashed curve), and the seismic$_{1}$B$_{31}$ 
(thin dot-dashed curve) models.} }
\end{figure*}

\begin{figure*}[pht*]
\epsscale{0.8}
\plotone{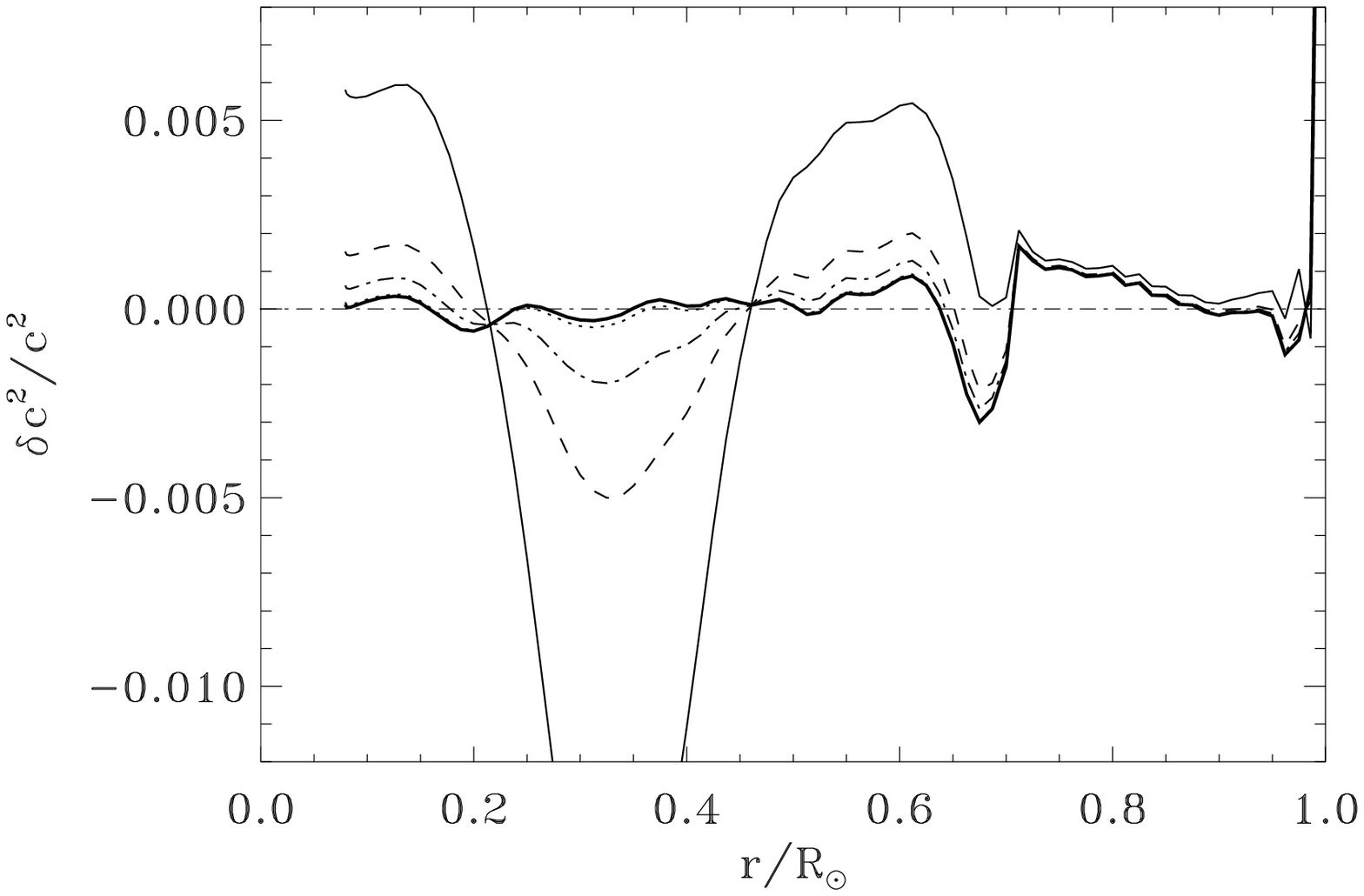}
\plotone{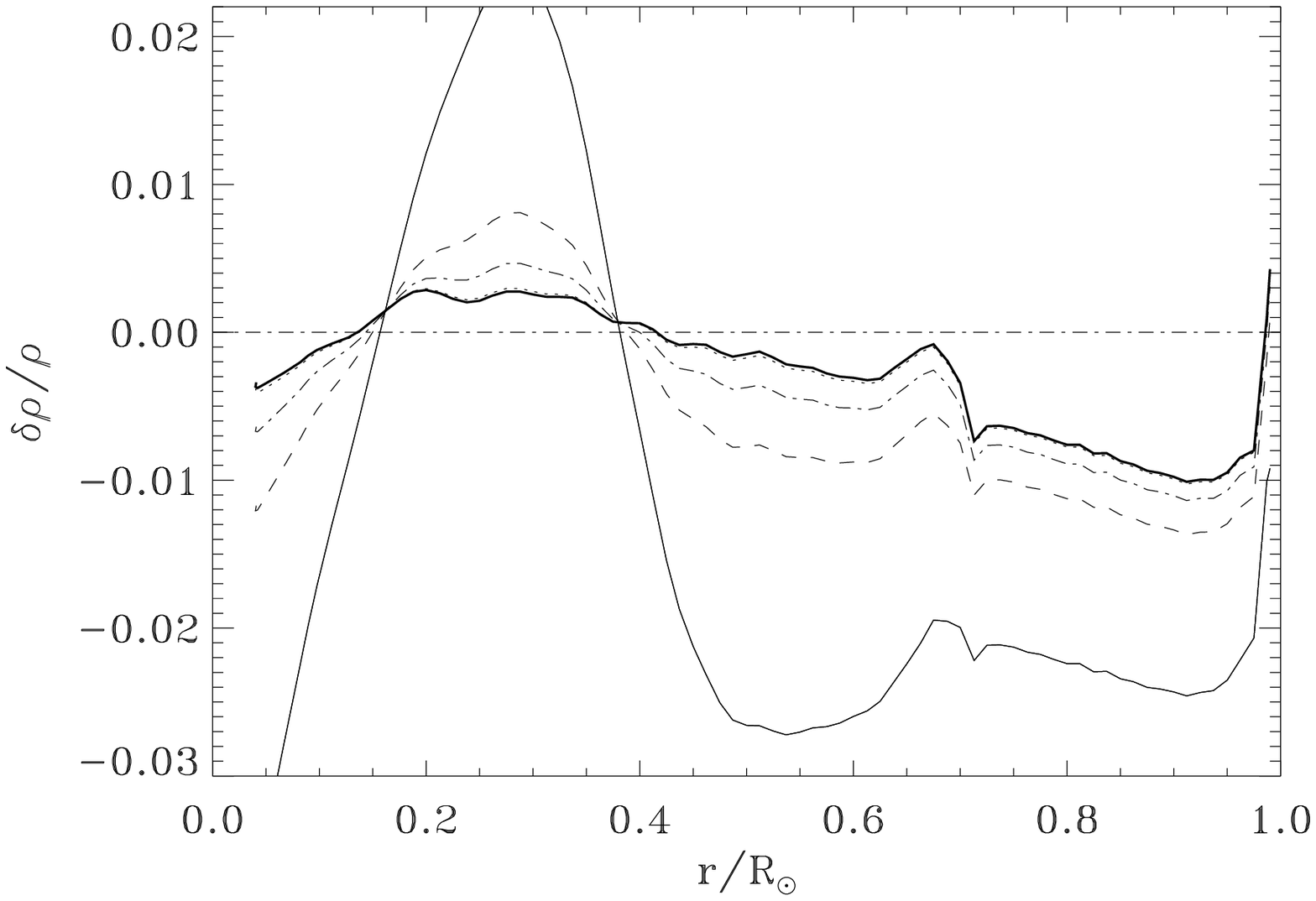}
\caption{\label{sismique1B1} {\it Upper figure: difference in the square of the sound speed 
between the Sun and the seismic$_{1}$B$_{1}$ model (thin plain curve). Same 
figure for the seismic$_{1}$B$_{11}$ model (dashed curve), the 
seismic$_{1}$B$_{12}$ model (dot-dashed curve), and the seismic$_{1}$B$_{13}$ 
model (dotted curve). The usual profile for the seismic$_{1}$ model has also 
been drawn (thick plain curve). Lower figure: idem for the density. } }
\end{figure*}

Since all the physical quantities considered here are radial quantities, we 
average the magnetic pressure $P_{mag}(r,\theta)$ over $\theta$.

\subsection{Impact of the Magnetic Pressure on the Solar Model: $c_{s}$, $\rho$, 
and the Neutrino Fluxes}

The three different magnetic pressure profiles are drawn on Figure \ref{mag}.
The addition of magnetic pressure in the radiative zone following the 
seismic$_{1}$B$_{1}$ model induces a great change in the thermodynamic 
quantities ($P$,$T$...), especially in the $c_{s}$ and $\rho$ profiles (see \mbox{Fig.} 
\ref{pressionB1} \& \ref{sismique1B1}). On the contrary, the 
impact of a field 10 times smaller following the seismic$_{1}$B$_{13}$ model is 
minuscule (see \mbox{Fig.} \ref{pressionB1} and \ref{sismique1B1}).
The precision we have on the solar sound speed and density rules out a magnetic 
field with such a profile and an intensity as large as $B_{0}=10^{4}$ T. 
Actually, we can put an upper limit for a (toroidal) magnetic field in the 
radiative zone of about $3 \times 10^{3}$ T. As soon as $B_{0} \le 10^{3}$ T, 
$c_{s}$ and $\rho$ are not sensitive enough to $P_{mag}$ and we cannot draw any 
conclusion about the likelihood of a field like the one of the 
seismic$_{1}$B$_{13}$ model. We show in Table 1 the impact on the neutrino production of the seismic$_{1}$B$_{11}$ model. The fluxes are slightly larger than the ones deduced from the adjustment in the physics of the seismic models.
The result obtained for the $\mathrm{^8B}$ flux: $5.23 \times 10^6 \mathrm{cm^{-2} s^{-1}}$ remains in agreement 
with the present SNO result.

We face the same problem when adding $P_{mag}$ at the BCZ and in the upper 
layers (see \mbox{Fig.} \ref{pressionB1}). It is even more difficult to draw any 
conclusions: the sound speed and density are not sensitive enough to such 
modifications. With the current accuracy we have on $c_{s}$ and $\rho$, it is 
only possible to state that a field in the tachocline can reach an amplitude as 
large as $50$ T without perturbing the sound speed profile. We conclude the same 
for a toroidal field anchored at $0.96 R_{\odot}$ and as large as $3$ T. Given 
the minuscule impact on $c_{s}$ and $\rho$, we did not draw the $\delta 
c_{s}^{2}/c_{s}^{2}$ and $d\rho/\rho$ profiles obtained with the fields at the 
BCZ and in the upper layers.

This section on the magnetic field confirms that $c_{s}$ and $\rho$ are only 
sensitive to the $\beta$ ratio. With the different models we computed, each one 
with a different magnetic field profile and/or intensity, we can conclude that 
only the large scale fields with $\beta^{-1}$ larger than, at least, $\simeq 3 
\times 10^{-4}$, impact on the sound speed and density profiles. The firm 
conclusion to draw is that an intensity as large as $3 \times 10^{3}$ T can be 
ruled out for a in-radiative-zone toroidal field. Despite the great accuracy we 
reached on this quantity, the sound speed is not suited to the determination of 
the large scale magnetic features of the Sun: many physical processes $c_{s}$ is 
quite sensitive to, are still affected by large error bars, and a potential 
magnetic field looks like ``background noise'' compared to these processes. The 
same conclusion is valid for the density profile. Yet the use of these two 
quantities might be more promising for constraining an upper-layer field, since 
the ``weakness'' of $P_{gas}$ near the solar surface makes $c_{s}$ more reactive 
to weaker fields. Unfortunately, this part of the Sun is also badly modeled 
due to the lack of turbulence in the present stellar equations.

Concerning the neutrino puzzle, none of the reasonable models including a central magnetic field
(as models seismic$_1$B$_{11}$, seismic$_1$B$_{12}$, \& seismic$_1$B$_{13}$) greatly modify the neutrino 
flux predictions, except the ruled out seismic$_{1}$B$_{1}$ model (it increases 
the $\mathrm{^{8}B}$ neutrino flux by more than $20\%$). With the upper bounds 
we have for the large scale magnetic field, it seems that this field has no 
impact on the neutrino emission, or only a very slight one.

\subsection{About the Relation Between Magnetic Field and the Neutrino Transport}

So far, we just considered mean static fields. We concluded that these fields probably
 impact only slightly on the neutrino production. However, the solar magnetic field is expected to be highly 
variable, and may locally be of large intensity: in the convective zone, the field is expected 
to be concentrated in flux tubes. Actually, the very idea of a mean field seems to be 
fallacious, at least in the external part. A variable and local magnetic field could act on the neutrino transport, since this field is expected to be by far larger than an hypothetic mean field.

We know from our previous analyses that a field with $P_{mag}/P_{gas}= 10^{-4}$ 
changes the gas pressure by at most $\delta P/P = -1.65 \times 10^{-4}$. 
It changes the sound speed by $\delta c_{s}^{2}/c_{s}^{2} = -0.74 \times 10^{-4}$, 
and the density by $\delta \rho / \rho \simeq -1  \times 10^{-4}$ 
(the minus sign means that when you add a magnetic field, 
you reduce the density and the sound speed). 
Of course, a change in the density means a change in the electron number 
and neutron number densities. Depending on the flux tube magnetic profile, 
and on the intensity of this field, such changes can impact on the neutrino 
transport: the appearance of a flux tube on the neutrino path may flip 
the neutrino spin, the sudden change in the electron and neutron number 
densities may act on the neutrino oscillations...

To correctly account for all these phenomena, 
we will have to use the results of magneto-hydro-dynamic simulations of the solar magnetic tubes.
 Thus we will get access to the way the electron and neutron number densities vary inside the tube, 
 and the magnetic profile of this one.
With a 1D stellar code, we cannot reproduce the complex magnetic structure of the Sun, but we can 
begin to work out some estimates, very near the surface where the local magnetic field is known and dominates the gas pressure.
We are ready to begin this study to estimate the impact on neutrino oscillations. 

\section{CONCLUSION}

Thanks to the most recent seismic data from the GOLF and MDI instruments aboard 
SoHO, precise sound speed and density profiles were derived down to 0.07 R$_\odot$,
that enabled us to 
build solar seismic models. The main goal was to obtain models in total 
agreement on the sound speed with the real Sun in the regions 
where no dynamic effects have been observed yet. The two first seismic models proposed
perfectly fulfill this requirement below $0.6 \, 
R_{\odot}$: it seems that further improvements in solar modelling require 3D 
codes to correctly account for the dynamic processes in the Sun. 

With the 
seismic models, we derive frequencies for acoustic and gravity modes and neutrino emitted fluxes: these fluxes are no more the 
results of purely theoretical considerations, but they also include seismic data. 
They are known with a better precision. Recent releases from the SNO 
collaboration announce a $\mathrm{^{8}B}$ neutrino flux received on Earth very 
close to the one predicted here. Of course, this is a 
great breakthrough in the understanding of the neutrino puzzle.
The seismic models considered so far are the result of long years of 
improvements in both the seismic data and the solar physics.
The sound speed is now quite sensitive to the $p-p$ reaction rate, the 
metallicity, the opacities, the solar age and even to some magnetic fields. This 
is a very powerful tool that might be useful to constrain the large scale solar 
magnetic field, provided that some further improvements are realized on the 
other physical parameters that have a large impact on $c_{s}$. It seems that the 
density profile should be closely monitored since it is more reactive than 
$c_{s}$ to the changes in several physical processes. The present density profile recently derived 
needs to be confirmed by an enhanced number of low frequency modes.
An important feature of the seismic$_{1}$ model is to favor a slight increase of 
the $p-p$ reaction rate by $\approx 1\%$ and the initial metallicity by about 
$3.5\%$. Actually, it is rather difficult to choose between an increase of the 
metallicity and a change of both the opacities and the metallicity. It also 
seems that the models must be calibrated at a radius smaller than the usual one 
(by $50-125$ km, depending on the atmosphere model), to improve the agreement 
with the Sun in the upper layers. This proposition needs to be cautiously considered, since a complex physics 
is present in the external layers, including turbulence and magnetic field.
Finally, the solar age $4.6$ Gyr is a good 
compromise between an older Sun favored by the sound speed profile and a younger 
one favored by the density profile. However, the density strongly disfavors an 
increase of $t_{\odot}$.

We put constraints on the magnetic field in the solar core and show that the fields simulated here do 
not deteriorate the agreement with the SNO observations, supposing three neutrino flavors.
We compute the different quantities useful to deduce the neutrino oscillation parameters, in the framework
of classical models. We also begin to estimate to what extent these classical models are representative of the real Sun.
Many progresses are still needed in the solar models ---more specifically for 
the upper layers--- but the high quality of the seismic data combined with the 
improvement in the physics make us already achieve a good result in solar 
modelling: the seismic$_{1}$ model seems very 
close to the real Sun, and the seismic$_1$B$_{11}$ model is also interesting to consider due to the presence of magnetic pressure.
 Contrarily to the usual seismic models, ours were 
obtained with a classical stellar evolution code. Nevertheless, as previously mentioned, 
as far as the very internal rotation profile is not included in such a study, 
new surprises may appear that invalidate the classical approach as it does for the solar convective zone.

\end{document}